\documentclass[preprint]{elsarticle}
\usepackage[british]{babel}
\usepackage[utf8]{inputenc}
\usepackage{graphics}
\usepackage{graphicx}
\usepackage{amsmath,amssymb,esint}
\usepackage{lineno}
\usepackage{multirow}
\usepackage{subfig}
\usepackage{natbib}
\usepackage{eucal}
\usepackage{mathtools}
\usepackage{placeins}
\usepackage{enumerate}
\usepackage[usenames,dvipsnames]{xcolor}
\usepackage[colorlinks]{hyperref}
\usepackage{setspace}
\usepackage{tikz}
\usepackage[nice]{nicefrac}
\usepackage{wasysym}
\usepackage{booktabs}
\usepackage{adjustbox}
\usepackage{array}	
\usepackage{color, colortbl}
\usepackage{pdfpages}
\usepackage{fullpage}
\usepackage{algorithm2e}[ruled]

\graphicspath{{./}{Figures/}}
\DeclareFontEncoding{LS1}{}{}
\DeclareFontSubstitution{LS1}{stix}{m}{n}
\DeclareMathAlphabet{\mathscr}{LS1}{stixscr}{m}{n}
\SetMathAlphabet{\mathscr}{bold}{LS1}{stixscr}{b}{n}

\usetikzlibrary{calc}
\usetikzlibrary{shapes.geometric, arrows}
\usetikzlibrary{arrows.meta}
\tikzset{>={Latex[width=1mm,length=1mm]}}

\tikzstyle{process} = [rectangle, minimum width=2.5cm, minimum height=1cm, text
centered, text width = 3cm, draw=black]
\tikzstyle{decision} = [diamond, minimum width=2.5cm, minimum height=1cm,
aspect=2,inner sep=-0.5ex,text centered, text width = 2.5cm, draw=black]
\tikzstyle{arrow} = [thick,->,>=stealth]
\tikzstyle{line} = [draw, -latex']

\biboptions{sort&compress}

\newcommand{\vect}[1]{\boldsymbol{#1}}

\newcommand\Rep{$\text{Re}_\textup{p}$}
\newcommand\cd{$C_\textup{D}$}
\newcommand\cdp{$C_{\textup{D},\parallel}$}
\newcommand\cdpp{$C_{\textup{D},\perp}$}
\newcommand\cdpl{$C_{\textup{D},\parallel}^{\text{\tiny{$\tilde{L}$}}}$}
\newcommand\cdl{$C_{\textup{D}}^{\text{\tiny{$\tilde{L}$}}}$}

\newcommand\cl{$C_\textup{L}$}

\newcommand\clpl{$C_{\textup{L},\parallel}^{\text{\tiny{$\tilde{L}$}}}$}
\newcommand\clpb{$C_{\textup{L},\parallel}^{\text{\tiny{${b}$}}}$}
\newcommand\cll{$C_{\textup{L}}^{\text{\tiny{$\tilde{L}$}}}$}

\newcommand\ct{$C_\textup{T}$}
\newcommand\ctp{$C_{\textup{T},\parallel}$}

\newcommand\ctpl{$C_{\textup{T},\parallel}^{\text{\tiny{$\tilde{L}$}}}$}
\newcommand\ctpb{$C_{\textup{T},\parallel}^{\text{\tiny{${b}$}}}$}
\newcommand\ctplp{$C_{\textup{T},\perp}^{\text{\tiny{$\tilde{L}$}}}$}
\newcommand\ctl{$C_{\textup{T}}^{\text{\tiny{$\tilde{L}$}}}$}

\makeatletter
\def\ps@pprintTitle{%
  \let\@oddhead\@empty
  \let\@evenhead\@empty
  \let\@oddfoot\@empty
  \let\@evenfoot\@oddfoot
}
\makeatother

\begin{document}
\begin{frontmatter}

\title{Drag, lift, and torque correlations for axi-symmetric rod-like non-spherical particles in linear wall-bounded shear flow}%

\author{Victor Ch\'{e}ron}
\author{Berend van Wachem\corref{cor1}}
\cortext[cor1]{Corresponding author}
\ead{berend.vanwachem@ovgu.de}
\address{Lehrstuhl f\"ur Mechanische Verfahrenstechnik, Otto-von-Guericke-Universit\"at Magdeburg, \mbox{Universit\"atsplatz 2, 39106 Magdeburg, Germany}}

\begin{abstract}
    This paper presents novel correlations to predict the drag, lift, and torque coefficients of axi-symmetric non-spherical rod-like particles in a wall-bounded linear shear flow. The particle position and orientation relative to the wall are varied to systematically investigate the influence of the wall on the hydrodynamic forces.
    The newly derived correlations for drag, lift, and torque on the particle depend on various parameters, including the particle Reynolds number, the orientation angle between the major axis of the particle and the main local flow direction, the aspect ratio of the particle, and the dimensionless distance from the particle centre to the wall.
    The impact of the wall on the hydrodynamic forces is accounted for as a function of a multiplication factor on the drag force in case of locally uniform flow, and an additional force contribution for the lift and the torque, modifying the resultant forces experienced by a particle in a locally uniform flow.
    The changes in the hydrodynamic forces prove to be substantial, emphasizing the necessity of accounting for wall effects across all particle types and flow conditions investigated in this study.
    The coefficients of the correlations are determined through a fitting process utilizing the data generated from our previous study on the interaction forces between a locally uniform flow and an axi-symmetric non-spherical rod-like particles, as well as from data of novel direct numerical simulations (DNS) performed in this work of flow past axi-symmetric rod-like particles near a wall. The proposed correlations exhibit a good agreement compared to the DNS results, with median errors of 2.89\%, 5.37\%, and 11.00\%, and correlation coefficients of $0.99$, $0.99$, and $0.96$ for the correlations accounting for changes in drag, lift, and torque coefficients due to the wall-bounded linear shear flow profile, respectively. These correlations can be used in large-scale simulations 
    using an Eulerian-Lagrangian or a CFD/DEM framework to predict the behaviour of axi-symmetric rod-like non-spherical particles in wall-bounded flows, shear flows, as well as uniform flows.
\end{abstract}

\begin{keyword}
Non-spherical particles \sep
Wall-bounded linear shear flow \sep
Drag, lift, and torque coefficients

\end{keyword}


\end{frontmatter}


\section{Introduction}

Understanding the dynamics of particle-laden flows is of primary importance due to their prevalence in environmental and industrial processes, such as river sediments transport, pneumatic conveying, and particle fluidization.
The complexity of these flows, attributed to the multitude of time and length scales involved, poses a challenge for reliable and cost-effective predictions. Numerical simulations have emerged as an effective tool to study particle-laden flows, offering a comprehensive and efficient means of investigation~\citep{Harting2014}.
Among the several numerical frameworks available~\citep{Subramaniam2022}, the Eulerian-Lagrangian (EL), CFD/DEM, or Lagrangian particle tracking (LPT) framework is often employed~\citep{Hilton2011,Li2013c,Mallouppas2013b,Zhao2013a,vanWachem2015,Kuerten2016,Fitzgerald2021,Mema2021}.
In this framework, analytical, semi-empirical, or empirical correlations are used to determine the interaction forces between fluid and particles, serving as momentum sources in the fluid momentum equations. In the EL framework, the accuracy of the particle-laden flow simulations heavily relies on the accuracy of the correlations predicting the interaction forces between the local fluid properties and the specific particle.
For spherical particles, several correlations exist, and include the effects of the fluid flow regime~\citep{Schiller1933,Clift1971}, the solid volume fraction of the particle assembly~\citep{Tenneti2011,Tang2015,Akiki2017,Hardy2022,vanWachem2023}, or the velocity profile of the fluid flow at the location of the particle~\citep{Kurose1999,Bagchi2002,Zeng2009}. 
These correlations are successfully employed in the EL framework for studying large-scale particle-laden flows applications, such as spherical particle fluidization~\citep{Fries2011,Gupta2016,Liu2019}, or their transport in turbulent pipe or channel flows~\citep{Picano2009,Marchioli2010,Mallouppas2013b}, to name just a few.\\

For non-spherical particles, however, an accurate correlation to predict the interaction forces between the local fluid flow and the particle is more complex to derive, since the shape of the particle, and the orientation of the particle with respect to the main local fluid flow direction, play a crucial role~\citep{Chhabra1999,Michaelides2023}.
For axi-symmetric particles, several correlations consider the shape of the particle, the orientation angle, along with the fluid flow regime, in the prediction of the interaction forces between non-spherical particles and the local fluid force.
In the viscous regime, where inertial effects can be neglected,~\citet{Jeffery1922}, and later on~\citet{Brenner1963}, derive analytical correlations to predict the hydrodynamic torque coefficient, in this work simply referred to as torque coefficient, and the drag and lift coefficients of a particle of varying elongation and orientation.
These correlations are used to study axi-symmetric particles transport, for instance in turbulent channel fluid flows~\citep{Marchioli2010,Zhao2013a}, or their deposition on a wall~\citep{Fan1995,Zhang2001}, to name a few applications.\\

To include inertia effects in the original correlations of~\citet{Brenner1963}, several numerical studies have emerged employing direct numerical simulations (DNS) to study interaction forces for a fixed isolated non-spherical particles subject to a uniform flow~\citep{Holzer2008,Zastawny2012c,Sanjeevi2018}.
This numerical tool allows for the easy modification of parameters such as the particle shape, fluid flow regime, or orientation angle between the main local fluid flow direction and the particle orientation, all while maintaining a detailed description of the particle-fluid interactions.
The particle-resolved DNSs (PR-DNS) yield an accurate calculation of the fluid stress over the surface of an isolated particle, enabling the derivation of accurate empirical correlations to model the drag, lift and torques coefficients~\citep{Zastawny2012c,Sanjeevi2018}.
While the aforementioned investigations derive correlations to model the fluid-particle interaction forces for a unique particle shape, the shape of the particle can also be included as a variable in the prediction of the forces and torques coefficients~\citep{Ouchene2015,Ouchene2016,Frohlich2020,Sanjeevi2022,Feng2023,Cheron2024}. 
Similarly to the analytical correlations of~\citet{Jeffery1922} and ~\citet{Brenner1963}, the correlations to predict the forces and torques coefficients at varying fluid flow regimes are also successfully used in EL simulations~\citep{Zhong2016}. For example, in studies focusing on non-spherical particle fluidization~\citep{Mahajan2018,Fitzpatrick2021,Mema2021,Atxutegi2021}, or non-spherical particle transport~\citep{Hilton2011,vanWachem2015}.\\

Although the correlations predicting the drag, lift, and torque coefficients in case of locally uniform flow at the location of the particle are used to study numerous particle-laden flow engineering applications, employing these correlations inevitably leads to neglecting variations in the fluid velocity profile.
Thus, relying on such hypotheses hampers the accuracy of the transport of the particles when the fluid flow is not uniform.
To account for the change in the hydrodynamic forces and torque coefficients in case of non-uniform flows, novel correlations are derived to model the fluid-particles interaction forces and torque for an unbounded linear shear flow past an isolated axi-symmetric rod-like particle based on a series of PR-DNS~\citep{Cheron2024}.
The results of their study show a significant change in the forces and torque coefficients compared to locally uniform flow, especially for the most elongated particles.
Understanding the effect of a wall on the interaction forces near the wall is explored in the studies by~\citet{Bhagat2022} and~\citet{Fillingham2021}.
In the latter, an empirical correlation is derived to predict the drag, lift, and torque coefficients for a prolate particle fixed to a wall.\\

The correlations of~\citet{Fillingham2021} enable the accurate calculation of the interaction forces between the fluid and a non-spherical particle fixed to a wall.
However,~\citet{Zhao2015} show that the preferential concentration of non-spherical inertial particles in turbulent channel flows is at a few distances from the wall.
Thus, an empirical correlation to model the transport of the non-spherical particles at distances ranging from far from a wall to close-contact with a wall in the EL framework is necessary to perform accurate particle-laden flow simulations with wall boundaries.
To propose such correlations, our study focuses on understanding the interactions between a fluid governed by a wall-bounded linear shear flow and a non-spherical fixed particle. We conduct a series of PR-DNS, systematically varying parameters such as the aspect ratio of the particle, the orientation angle between the main semi-major axis of the particle and the main local fluid flow direction, the distance between the particle and the wall, and the flow regime.
In this work, we vary the orientation angle with respect to the stream-wise and wall-normal direction, aligning with the observation that the orientation angle of non-spherical particles in wall-bounded turbulent flows predominantly varies in this plane~\citep{Zhang2001,Mortensen2008,Marchioli2010,Zhao2015}.
The three non-spherical axi-symmetric rod-like particles considered within this work are shown in the table~\ref{table:fibersshapes}.\\

From the results of the PR-DNS, empirical correlations for the drag and lift forces as well as the torque are derived.
The simulation results of~\citet{Zeng2009}, who investigate the change in the force coefficients for a spherical particle at varying distance from a wall, indicate that the influence of the wall boundary reduces as a function of the distance to the wall.
Far from the wall, the force coefficients of the spherical particle reduce to the forces coefficients in the case of uniform flow.
Thus, the correlations presented in this work are built upon the correlations derived by~\citet{Cheron2024} for unbounded shear flow, which are also used as a limit, to predict the hydrodynamic forces for particles far from a wall.
The novel correlations are valid for varying fluid flow regimes, orientation angles, aspect ratios, and varying distances from the wall.
The correlations can be applied to predict the behaviour of axi-symmetric rod-like non-spherical particles at any distance from a wall, including within the viscous boundary layer of wall-bounded fluid flow configurations, in the EL framework.\\
\begin{table}
\begin{tabular}{rccc}
       &
      \includegraphics[width=0.25\columnwidth, trim={400 20 400 20},clip]{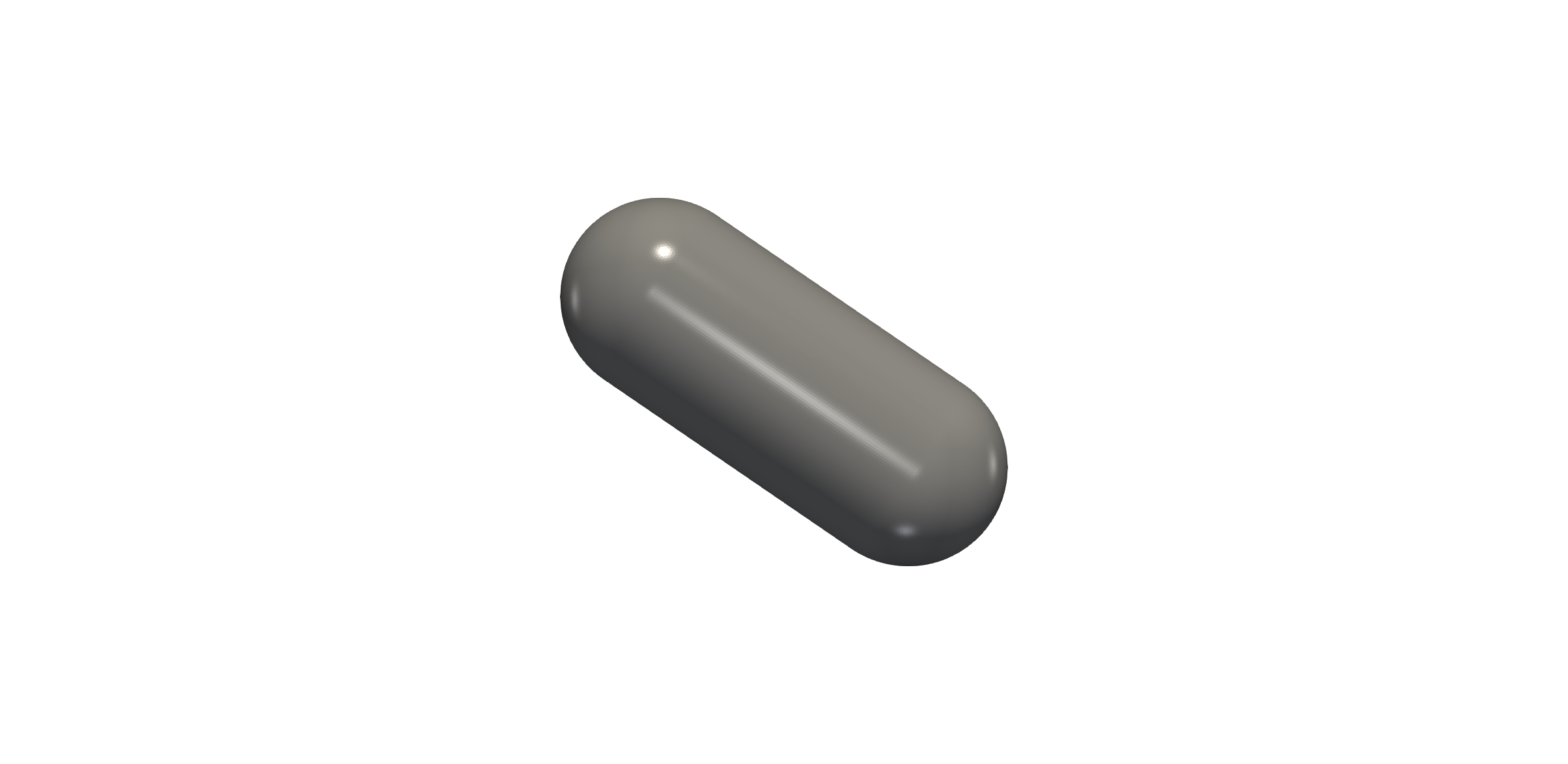} &
      \includegraphics[width=0.25\columnwidth, trim={400 20 400 20},clip]{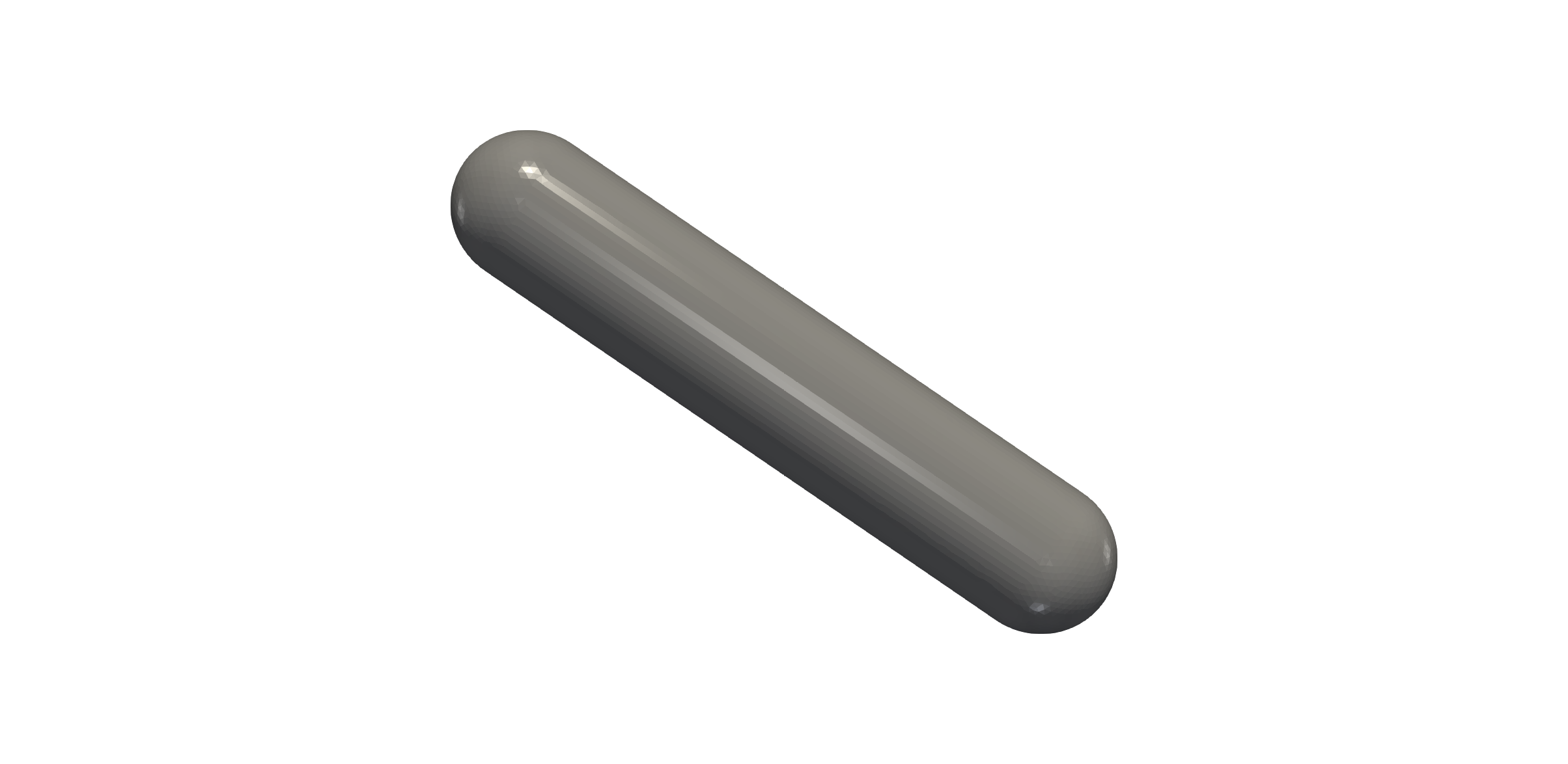} &
      \includegraphics[width=0.25\columnwidth, trim={400 10 400 10},clip]{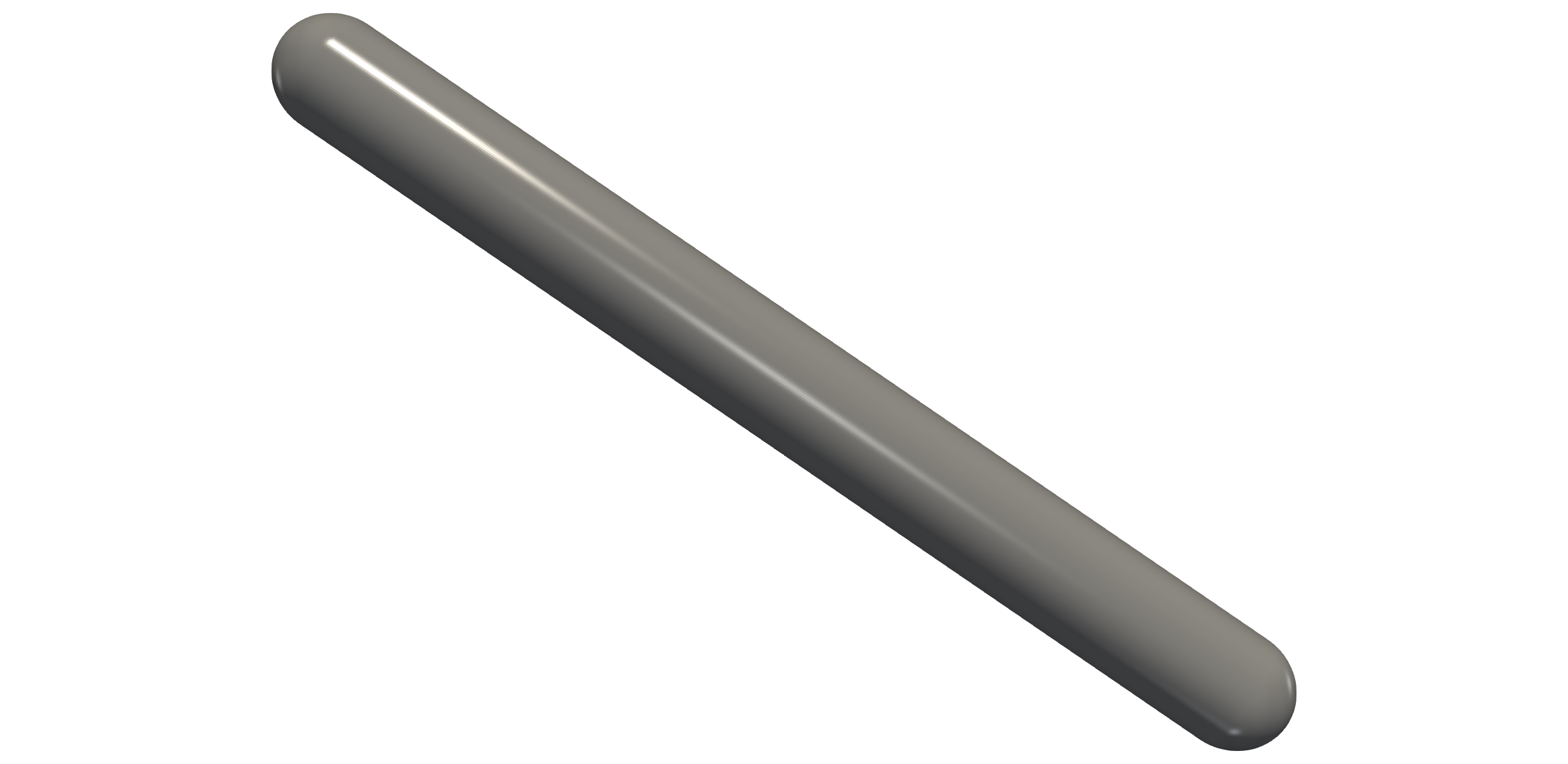} \\
      \small Aspect ratio $\alpha=\frac{a}{b}$ & 2.5 & 5 & 10 \\

      \small Sphericity $\Phi=\frac{A_\textup{S}}{A_\textup{p}}$ & 0.878 & 0.737 & 0.596 \\
\end{tabular}
     \caption{The overview of the three axi-symmetric rod-like non-spherical particles considered in this paper and their shape coefficients. $\Phi$ is the ratio of the surface of the volume-based-equivalent sphere, $A_\textup{S}$, and the surface of the particle, $A_\textup{p}$. $\alpha$ is the ratio between the semi-major, $a$, and semi-minor $b$, axes of the particle.
     }\label{table:fibersshapes}
\end{table}\\

This paper is organised as follows: Section~\ref{sec:forcesonparticle} describes the flow configuration and the parameters varied in this study, Section~\ref{sec:numericalframework} presents the numerical method employed to perform the PR-DNS. Section~\ref{sec:description} describes the numerical configuration to perform the PR-DNS. The results of the PR-DNS, and the correlations to predict the fluid-particles interaction forces and torque coefficients between non-spherical axi-symmetric rod-like particles and these typical flows, are discussed in Section~\ref{sec:results}. Finally, conclusions of this work are presented in Section~\ref{sec:conclusions}.

\section{Simulation setup\label{sec:forcesonparticle}}

\begin{figure}[htbp!]
\centering
\includegraphics[width=0.35\columnwidth]{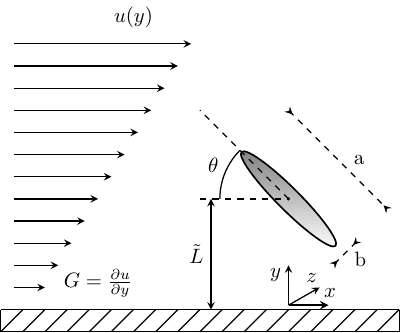}
    \caption{Flow configuration of the axi-symmetric particle in a wall-bounded linear shear flow study.}\label{fig:flowconfiguration}
\end{figure}

The flow configuration consists of a fixed axi-symmetric particle subject to a linear shear flow bounded by a smooth flat wall.
The configuration is shown in figure~\ref{fig:flowconfiguration}; the frame of reference is attached to the bottom wall, with the $x$ and $z$ origins that coincide with the centre of mass of the particle.
The $x$ co-ordinate represents the stream-wise direction, the $y$ co-ordinate is the wall-normal direction and $z$ co-ordinate is the span-wise direction.
Similar configurations have been thoroughly investigated to predict force coefficients for spherical particles~\citep{Zeng2009}, and axi-symmetric particle fixed to the wall~\citep{Fillingham2021}.
The orientation of the particle between its major axis and the main local fluid direction is characterized by the orientation angle $\theta$, which in this work is varied only in the $x$-$y$ plane.
In this work, an orientation angle of $\theta = 0^{o}$, or $\theta = 180^{o}$, indicates that the semi-major axis of the particle is parallel to the wall, and, $\theta = 90^{o}$ indicates that the semi-major axis of the particle is perpendicular to the wall. 
In the study of the modelling of the interaction forces between unbounded linear shear flow and non-spherical particle of~\citet{Cheron2024}, it is shown that the symmetric behaviour of the interaction forces between the local fluid flow and the particle about the orientation angle $\theta = 90^\textup{o}$ is no longer achieved.
Thus, the orientation angle $\theta$ is varied in the range $\theta = 0^\textup{o}$ to $180^\textup{o}$.\\

The shape of the particle is characterized by its aspect ratio, $\alpha$, which is defined as the ratio between the major and the minor axes of revolution of the particle, $a$ and $b$,
\begin{equation}
\alpha = \frac{a}{b}\, ,\quad\text{ with }\quad b\leq a\, .
\end{equation}
The particles studied in this work have an aspect ratio of $\alpha=2.5$, 5 and 10.\\

The distance between the non-spherical particle and the wall can be characterized by the distance between the centre of the particle and the wall, or by the shortest distance between any point at the surface of the particle to the wall.
As mentioned in the review of~\citet{Ma2022}, in which several contact detection algorithms for non-spherical particles are discussed, the accurate computation of the shortest distance between a non-spherical particle and a wall for the use of the correlations in the EL framework may lead to a subsequent increase in computational cost.
Thus, in this work we use the distance between the centre of the particle and the closest distance to the wall, which is also the origin of the reference frame.
This distance changes only when the distance between the centre of the particle and the wall in the direction of the $y$ co-ordinate is changed, and is given in dimensionless form as $\tilde{L} = L/D_\text{eq}$, with $L$ the dimensional distance from the wall to the centre of the particle, and $D_\text{eq}$, the volume-based-equivalent diameter of the non-spherical particle given by
\begin{equation}
D_\text{eq} = 2 \left(\frac{3 V_\textup{p}}{4}\right)^{1/3}\, , 
\end{equation}
where $V_\textup{p}$ is the volume of the specific particle.

\begin{figure}[htbp!]
    \centering
    \includegraphics[width=0.65\columnwidth, trim={15 130 20 130}, clip]{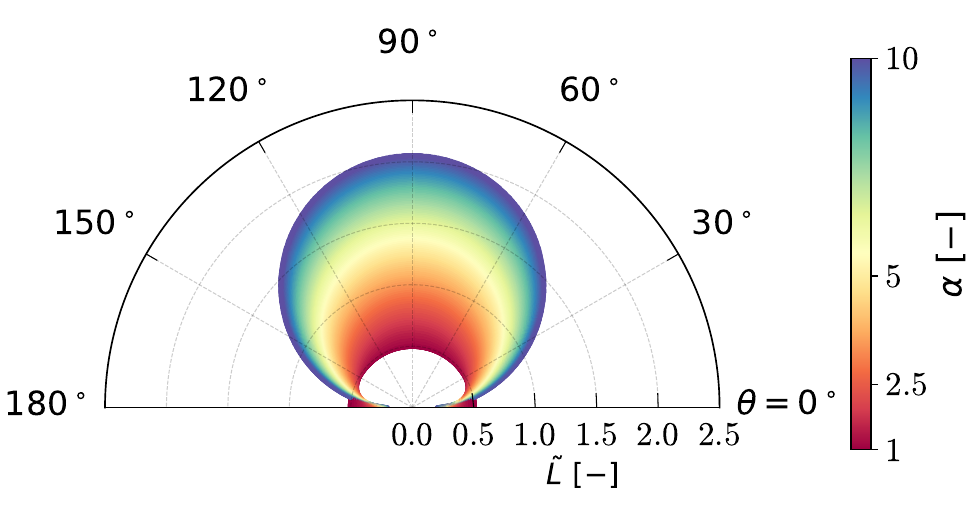}
        \caption{Shortest dimensionless distance to the wall, $\tilde{L}$, as a function of the orientation angle, $\theta$, for particles with aspect ratios ranging from $\alpha = 1$ to $\alpha = 10$. }\label{fig:CollisionMap}
\end{figure}

In this work, the dimensionless distance from the centre of the particle to the wall, $\tilde{L}$, varies in the range of $\tilde{L}=4$ to $b$, where $\tilde{L}=b$ represents a particle fixed to the wall.
Following the methodology of~\citet{Zeng2009} for closing their model to predict the hydrodynamic forces acting on a spherical particle in a wall boundary layer, we use the results from uniform flow cases to represent conditions equivalent to an infinite distance from the wall, \textit{i.e.,} when the particle is outside the wall boundary layer. 
The hydrodynamic forces and torque on identical particles under uniform flow conditions, obtained from the PR-DNS of~\citet{Cheron2024}, are indicated by setting the distance $\tilde{L} \rightarrow \infty$. For correlation derivation purposes, this distance is specifically set to $\tilde{L}=50$.
It is worth mentioning that since the definition of the distance is independent of the orientation of the particle, it yields to several dimensionless distances where the particle intersects the wall.
For example, any particle with aspect ratio larger than $1$, at a wall distance $\tilde{L} = b$, with an orientation angle not equal to $\theta = 0^{o}$ or $180^{o}$, intersects with the wall.
A more general formulation discriminates these cases and provides the range of validity of the correlations near the wall, given by
\begin{equation}\label{eq:CollisionDistance}
    \begin{cases}
        \text{if the dimensionless distance is less than}\quad \tilde{L} \leq \nicefrac{b}{2} & \text{wall-particle ovelap,}\\
        \text{if the dimensionless distance is less than}\quad \nicefrac{\tilde{L}}{\cos{(90-\theta)}} \leq \nicefrac{a}{2} & \text{wall-particle overlap,}\\
        \text{else}\quad & \text{correlation is applicable,}\\
    \end{cases} 
\end{equation}
and is also illustrated in figure~\ref{fig:CollisionMap}.
\\

\citet{Zeng2009} and~\citet{Fillingham2021} characterize the flow regime based on the particle Reynolds number, given by
\begin{equation}
\text{Re}_\textup{p} = \frac{u_\textup{f@p} D_\text{eq}}{\nu_\textup{f}}\, ,
\end{equation}
with $\nu_\textup{f}$ the kinematic viscosity of the fluid, and $u_\textup{f@p}$ the conceptual undisturbed velocity at the location of the centre of the particle.
For a wall-bounded linear shear flow configuration, $u_\textup{f@p}$ is determined by the shear of the fluid flow and the distance to the wall.
Several possibilities exist to vary the particle based Reynolds number, and as we keep the velocity at the location of the particle constant, we adapt the kinematic viscosity of the fluid to vary the particle Reynolds number.
As the particle Reynolds number remains relatively low, the flow profile can be assumed linear~\citep{Zeng2009}.\\

The list of parameters and their range of variation which are varied to derive the empirical correlations for the drag,  lift, and torque coefficients are shown in table~\ref{table:matrix-simulation}.
For some configurations, such as for a particle of aspect ratio $\alpha = 10$ at a dimensionless distance to the wall of~$\tilde{L} = 1$, most of the orientation angles given in the table~\ref{table:matrix-simulation} cannot be considered, as the particle intersects the wall. Thus, additional orientation angles are considered to enable the derivation of the correlations to predict the drag, lift and torque coefficients.
These additional angles are given by the shortest interstice distance between the surface of the particle and the wall, see Eq.~\eqref{eq:CollisionDistance}.
This result in a total of 636 individual PR-DNSs to be performed to derive the novel correlations.
We have also incorporated data from other research works to establish the correct limits of the correlations proposed in our study. This includes scenarios beyond our current investigation scope, such as particles with an aspect ratio of 1 or particles situated in flows significantly distant from the wall.
\\

Finally, the drag and lift forces and torque coefficients predicted by the correlations for describing the interactions between non-spherical particles and the fluid flow are obtained in our framework by
\begin{equation}
C_\textup{D} = \frac{F_\textup{x}}{\frac{1}{2} \rho_\textup{f} \tilde{v}^2 \frac{\pi}{4} D_\textup{eq}^2}\, ,\quad
C_\textup{L} = \frac{F_\textup{y}}{\frac{1}{2} \rho_\textup{f} \tilde{v}^2 \frac{\pi}{4} D_\textup{eq}^2}\, ,\quad
C_\textup{T} = \frac{T_\textup{z}}{\frac{1}{2} \rho_\textup{f} \tilde{v}^2 \frac{\pi}{8} D_\textup{eq}^3}\, ,\quad
\end{equation}
with $\tilde{v}$ the relative velocity between the conceptual undisturbed velocity of the fluid at the location of the centre of the particle, $u_\textup{f@p}$, and the velocity of the particle.
Here $F_\textup{x}$, $F_\textup{y}$ and $T_\textup{z}$ are the dimensional forces and torque components for the drag force, the lift force and the torque about the span-wise co-ordinate of the simulation, respectively, and are directly obtained from the PR-DNS framework~\citep{Cheron2023a}.
The values of the remaining force and torque components are negligible, and are not discussed in this work.

\begin{table}
\centering
\begin{tabular}{ {l}|{l} }
 $\alpha$ $[-]$   & 2.5, 5, 10 \\
 \hline
 $\text{Re}_\textup{p}$ $[-]$ & 0.1 2, 10, 50, 100, 200, 300\\
\hline
 $\theta$ $[^\text{o}]$ & 0, 30, 45, 60, 90, 120, 135, 150\\
 \hline
 $\tilde{L}$ $[-]$  & $b$, $1.05b$, $1.1b$, $1.15b$, $1.25b$, $1.5b$, $2b$, $a$, 1, 2, 4\\

\end{tabular}
     \caption{Matrix of varying parameters for the simulation of the wall-bounded linear shear flow configuration.
     }\label{table:matrix-simulation}
\end{table}

\section{Numerical framework\label{sec:numericalframework}}

The PR-DNS of the wall-bounded linear shear flow past a fixed axi-symmetric particle is performed with the direct forcing smooth immersed boundary method (IBM)~\citep{Peskin1977}, using the direct-forcing formulation of~\citet{AbdolAzis2019}.
The IBM couples the modelling of the fluid domain through an Eulerian framework and the representation of the surface of the particle with a Lagrangian framework.
The Lagrangian framework consists of evenly spaced markers discretizing the surface of the particle.
The Eulerian framework is based on a finite-volume framework with collocated variable arrangement.
The incompressible fluid phase, subject to the Navier-Stokes equations, is solved with an implicit pressure-velocity coupling~\citep{Denner2014a}, and the source terms of the momentum equations are discretised to numerically balance the flow pressure gradient~\citep{Bartholomew2018}.\\

In the IBM, the Eulerian and Lagrangian frameworks are independent, and an interpolation/spreading strategy is used to transfer the Eulerian and Lagrangian fluid variables~\citep{Peskin2003}.
The interpolation and spreading compact supports are constructed through a mollified moving-least-squares algorithm~\citep{Bale2021}.
The size of the compact support determines the number of fluid cells used for the interpolation and the spreading of the fluid variables, in this work a five-point spline kernel function is used~\citep{Bao2016}.
For a particle lying on a wall-boundary, the interpolation and spreading of the fluid variables and source terms, respectively, exceed the computational domain.
Thus, the HyBM method of~\citet{Cheron2023a} is employed to restrain the compact support to inside the particle when a Lagrangian marker is near the boundary of the domain, typically for configurations with ${L} \rightarrow b$ at $\theta = 0^{o}$, or ${L} \rightarrow a$ at $\theta = 90^{o}$.
The spreading of the Lagrangian IBM source terms toward the Eulerian source terms of the fluid momentum equations is scaled by a relaxation factor~\citep{Zhou2021}.
This relaxation factor is based on stability condition criterion, and controls the rate at which the no-slip condition is reached as well the magnitude of the error of the no-slip.
The reader is referred to the work of~\citet{Cheron2023a} for the details of implementation and validation of the present IBM.\\

The terms of the Navier-Stokes equations are discretised as follows, a second-order Laplacian discretization scheme for the diffusion term, a second-order Minmod TVD scheme for the discretization of the advection term,
and a second order backward Euler temporal scheme for the transient terms~\citep{Denner2020}.
The time-step adapts for each time-step and is calculated based on the convection and viscosity criteria, as given in~\citet{Kang2000}, thus the most restrictive time-step is observed for the low Reynolds number simulations.
To precisely enforce the no-slip criteria on the particle surface in the IBM framework, a low CFL number is necessary, which is set to 0.05 in this study.
This ensures that the maximum error in the magnitude of the no-slip error at the surface of the particle is always lower, or equal to, 1\% of fluid flow penetration for all the PR-DNSs performed in this work.
Finally, the number of Lagrangian markers used to discretise the surface of the particle is determined by the fluid mesh cell length. By matching the distance between markers to the Eulerian fluid mesh cell length, a finer Eulerian fluid grid results in a higher number of markers.
The HyBM method has been used in several particle-laden flow studies, such as to study the interaction of a flow with monodispersed particles in an assembly~\citep{vanWachem2023}, to study the interaction of a flow with non-spherical isolated particle~\citep{Cheron2024}, or to study of the flow past a fixed packed bed reactor~\citep{Gorges2024,Patil2023}.

\section{Numerical configuration and domain independence study\label{sec:description}}

\begin{figure}[ht!]
    \centering
    \includegraphics[width=0.95\columnwidth, trim={0 0 0 0}, clip]{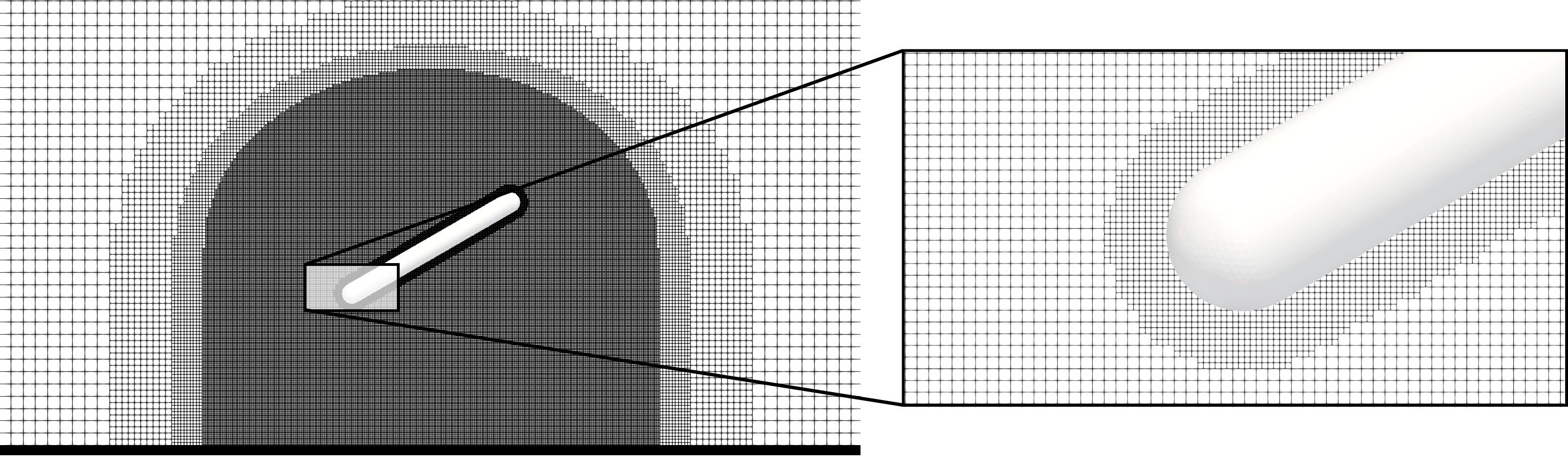}
    \caption{A cross-section of the mesh of one of the multi refinement levels fixed fluid mesh grid used to perform the PR-DNS of the present study. This specific configuration presents a particle of aspect ratio $\alpha=10$, a dimensionless distance of $\tilde{L} = 4$, and an orientation angle of~$\theta=30^{o}$.}
    \label{fig:mesh}
\end{figure}

The PR-DNS framework is employed to determine the drag, lift, and torque of the non-spherical particles shown in table~\ref{table:fibersshapes}.
A similar configuration has been studied for a spherical particle in our previous work~\citep{Cheron2023a}.
Thus, the same boundary conditions and numerical resolution are used in this work, and are summarized in table~\ref{table:boundaryconditions}.
For this specific case, the maximum level of refinement is set to $\Delta x = D_\textup{eq}/64$, with $\Delta x$ being the fluid mesh cell length.
This resolution is sufficient to provide accurate estimation for the forces experienced by a non-spherical particle subject to unbounded linear shear flow configurations~\citep{Cheron2024} as well as forces experienced by a spherical particle in wall-bounded linear shear flow configurations~\citep{Cheron2023a}.
Since the interest of the study lies in the force experienced by the particle only, the finest fluid mesh cell length is used only near the particle surface, using a fixed fluid mesh with multiple refinement levels.
Further away from the particle, the fluid mesh is coarsened to relax the required computational effort, except in the interstice distance between the wall and the particle, where a fine resolution of the fluid mesh is kept.
An example of a 2D cross-section of the fixed fluid mesh grid generated to perform the PR-DNS is shown in figure~\ref{fig:mesh} for a particle with aspect ratio $\alpha = 10$, orientation angle $\theta = 30^{o}$, dimensionless distance to the wall $\tilde{L} = 4$, with centre of co-ordinates of the particle located at $\vect{X}_\text{p} = [0, \tilde{L}, 0]$.\\
\begin{table}
    \centering
    \begin{tabular}{l l l l l l}
    $x^{-}$ & $x^{+}$ & $y^{-}$ & $y^{+}$ & $z^{-}$ & $z^{+}$ \\ 
     \hline\\
     $u_\textup{x} = G y$ & $\frac{\partial u_\textup{x}}{\partial x} = 0$ & $u_\textup{x} = 0$ & $u_\textup{x} = G L_y$ &  $u_\textup{x} = G y$ &  $u_\textup{x} = G y$\\[1ex]
     $u_\textup{y} = 0$ & $\frac{\partial u_\textup{y}}{\partial x} = 0$ & $u_\textup{y} = 0$ & $u_\textup{y} = 0$ &  $u_\textup{y} = 0$ &  $u_\textup{y} = 0$\\[1ex]
     $u_\textup{z} = 0$ & $\frac{\partial u_\textup{z}}{\partial x} = 0$ & $u_\textup{z} = 0$ & $u_\textup{z} = 0$ &  $u_\textup{z} = 0$ &  $u_\textup{z} = 0$\\[1ex]
     $\frac{\partial P}{\partial x} = 0$ & $P = 0$ & $\frac{\partial P}{\partial y} = 0$ & $\frac{\partial P}{\partial y} = 0$ & $\frac{\partial P}{\partial z} = 0$ &  $\frac{\partial P}{\partial z} = 0$\\
     
    \end{tabular}
         \caption{Boundary conditions for the three components of the fluid velocity vector $\mathbf{u}$ and the pressure $P$ to simulate the wall-bounded linear shear flow configuration, shown in figure~\ref{fig:flowconfiguration}. Here, $x$ is the stream-wise co-ordinate direction, $y$ is the wall-normal co-ordinate direction, which is 0 at the wall ($y^{-}$) and equal to the domain length at $y^{-}$, $y =  \text{L}_y$, and $z$ is the span-wise co-ordinate direction. $G$ represents the dimensional shearing of the wall-bounded flow.
         }\label{table:boundaryconditions}
\end{table}\\

To study the fluid flow past an isolated particle, ~\citet{Sanjeevi2018} mention that the computational domain must be sufficiently large for simulations with low particle Reynolds number, typically for $\text{Re}_\textup{p} = 0.1$.
At higher particle Reynolds number, typically starting at $\text{Re}_\textup{p} = 200$, the domain length in the wake of the particle should be kept large enough to prevent an influence of the outlet boundary.
Thus, a computational domain size independence study is performed for both particle Reynolds numbers in which the computational domain size is varied, while the fluid mesh cell length is kept constant, \textit{e.g.} the larger the computational domain size the more fluid cells are used to perform the PR-DNS.
The results of the computational domain size independence study at particle Reynolds number $\text{Re}_\textup{p} = 0.1$ provides the computational domain size for this flow regime, the computational domain size independence study at particle Reynolds number $\text{Re}_\textup{p} = 200$ provides the computational domain size for the remaining flow regimes, since they do not require a larger domain~\citep{Sanjeevi2018}.
For these simulations, the most restrictive configuration is considered, thus the dimensionless distance to the wall is set to $\tilde{L} = 4$,  the more elongated particle is considered, $\alpha = 10$, and two orientation angles are considered, $\theta = 0^{o}$ and $90^{o}$.
The computational domain size is given as a function of the volume-based-equivalent diameter, and the centre of co-ordinates of the particle, $\vect{X}_\text{p}$, is always set at $\vect{X}_\text{p} = [0, \tilde{L}, 0]$.\\

At particle Reynolds number $\text{Re}_\textup{p} = 0.1$, four computational domains, given in table~\ref{tab:domainsizestudyRep01}, are studied.
In the same table the results for the drag, lift and torque coefficients are given for all configurations.
The results obtained with the computational domain denoted by the abbreviation \Rep-01-C, as defined in table~\ref{tab:domainsizestudyRep01}, experience a change in the forces and torque coefficients lower than 1\% with respect to the reference, the largest computational domain, domain \Rep-01-D.
Thus, the simulations for particle Reynolds number  $\text{Re}_\textup{p} = 0.1$ are performed with the computational domain \Rep-01-C.\\

At particle Reynolds number $\text{Re}_\textup{p} = 200$, three computational domain sizes are studied. They are given in table~\ref{tab:domainsizestudyRep200}, along with the resulting drag, lift, and torque coefficients.
At this particle Reynolds number, for the two orientation angles, $\theta = 0^{o}$ and $90^{o}$, the domain size slightly influences the coefficients.
Thus, the simulations from particle Reynolds number $\text{Re}_\textup{p} = 1$ to $200$ are performed with the computational domain referred to as \Rep-200-B.

\begin{table}
\centering
\resizebox{\columnwidth}{!}{%
\begin{tabular}{l | c c c | c c c| c c c}
Domain & L$_{x}$ &  L$_{y}$ & L$_{z}$ & {$C_\textup{D}$} & {$C_\textup{L}$} & {$C_\textup{T}$}  & {$C_\textup{D}$} & {$C_\textup{L}$} & {$C_\textup{T}$}\\
& &   &  & $\theta = 0^{o}$ & $\theta = 0^{o}$ & $\theta = 0^{o}$ & $\theta = 90^{o}$ & $\theta = 90^{o}$ & $\theta = 90^{o}$\\
 \hline
 \hline
  \Rep-01-A & $[-24 D_\textup{eq}:30 D_\textup{eq}]$ & $[0 :30 D_\textup{eq}]$ & $[-13 D_\textup{eq}:13 D_\textup{eq}]$ & 343.53 & 4.04 & -25.74 & 499.80 & 4.63 & -479.15\\
  \Rep-01-B & $[-30 D_\textup{eq}:36 D_\textup{eq}]$ & $[0 :36 D_\textup{eq}]$ & $[-15 D_\textup{eq}:15 D_\textup{eq}]$ & 339.01 & 4.32 & -25.45 & 491.43 & 4.88 & -473.85\\
  \Rep-01-C & $[-36 D_\textup{eq}:46 D_\textup{eq}]$ & $[0 :40 D_\textup{eq}]$ & $[-18 D_\textup{eq}:18 D_\textup{eq}]$ & 337.75 & 4.47 & -25.31 & 488.88& 5.01 & -472.59\\
  \Rep-01-D & $[-40 D_\textup{eq}:50 D_\textup{eq}]$ & $[0 :44 D_\textup{eq}]$ & $[-20 D_\textup{eq}:20 D_\textup{eq}]$ & 337.26 & 4.49& -25.28 & 487.86& 5.04& -471.34\\

\end{tabular}%
}
     \caption{Computational domain size independence study for the flow regime Re$_\text{p} = 0.1$, dimensionless distance to the wall $\tilde{L} = 4$, and orientation angles {$\theta = 0^{o}$} and {$\theta = 90^{o}$}.}\label{tab:domainsizestudyRep01}
\end{table}

\begin{table}
\centering
\resizebox{\columnwidth}{!}{%
\begin{tabular}{l | c c c | c c c| c c c}
Domain & L$_{x}$ &  L$_{y}$ & L$_{z}$ & {$C_\textup{D}$} & {$C_\textup{L}$} & {$C_\textup{T}$}  & {$C_\textup{D}$} & {$C_\textup{L}$} & {$C_\textup{T}$}\\
 & &   &  & $\theta = 0^{o}$ & $\theta = 0^{o}$ & $\theta = 0^{o}$ & $\theta = 90^{o}$ & $\theta = 90^{o}$ & $\theta = 90^{o}$\\
 \hline
 \hline
 \Rep-200-A & $[-8 D_\textup{eq}:16 D_\textup{eq}]$ & $[0 :10 D_\textup{eq}]$ & $[-5 D_\textup{eq}:5 D_\textup{eq}]$ & 0.56 & $7.25 \times 10^{-3}$ & $2.41 \times 10^{-2}$ & 2.58 & $-4.27 \times 10^{-3}$ & -1.67\\
 \Rep-200-B & $[-12 D_\textup{eq}:20 D_\textup{eq}]$ & $[0 :14 D_\textup{eq}]$ & $[-7 D_\textup{eq}:7 D_\textup{eq}]$ & 0.56 & $7.31 \times 10^{-3}$ & $2.40 \times 10^{-2}$ & 2.56 & $-4.27 \times 10^{-3}$ & -1.68\\
 \Rep-200-C & $[-16 D_\textup{eq}:24 D_\textup{eq}]$ & $[0 :18 D_\textup{eq}]$ & $[-10 D_\textup{eq}:10 D_\textup{eq}]$ & 0.56 & $7.31 \times 10^{-3}$ & $2.40 \times 10^{-2}$ & 2.56 & $-4.28 \times 10^{-3}$ & -1.68\\
 
\end{tabular}%
}
     \caption{Computational domain size independence study for the flow regime Re$_\text{p} = 200$, dimensionless distance to the wall $\tilde{L} = 4$, and orientation angles {$\theta = 0^{o}$} and {$\theta = 90^{o}$}.}\label{tab:domainsizestudyRep200}
\end{table}

\section{Results\label{sec:results}}

To ensure time independent averaged results, the drag and lift forces and the torque are determined from the PR-DNSs after a sufficiently long physical time has passed since the start of the simulation.
The reference time is chosen based on previous work~\citep{Cheron2024}, and is set to $t_\text{ref} = \nicefrac{t \vert u_{f@p}\vert }{D_\text{eq}} = 200$, where $t$ is the physical time.
These data are used to derive novel correlations to predict the drag, lift, and torque coefficients as a function of the particle Reynolds number, \Rep, the orientation of the particle with respect to the main local flow direction, $\theta$, the aspect ratio of the particle, $\alpha$, and the dimensionless distance to the wall, $\tilde{L}$
In this study, the method for generating new correlations is based on symbolic regression algorithms, utilizing the PySR library~\citep{Cranmer2023}. The regression process employs an internal search mechanism, characterized by an evolve-simplify-optimize loop, to effectively refine empirical formulas. After identifying a set of potential expressions, their constants are optimized using curve fitting methods, particularly with the Levenberg-Marquardt algorithm from the SciPy library~\citep{Virtanen2020}.
The accuracy of the novel correlations is evaluated through the calculation of the coefficient of determination, $\mathcal{R}^{2}$, and median prediction error, ${\mathcal{E}}$.
The median prediction error is given by,
\begin{equation}
\mathcal{E} = \text{median} \left(\vert\vert f_{\text{model},i} - f_{i}\vert\vert/\vert\vert \max\left(f_{\text{model},i},f_{i}\right)\vert\vert\\\right)\, ,
\end{equation}
where $f_{i}$ represents the value of a given force or torque coefficient, for the $i-$th PR-DNS simulation, and $f_{\text{model},i}$ is the value predicted by the correlation.
The median prediction error is scaled by the maximum value of the prediction and the PR-DNS value, because the force coefficient can be equal to zero.
This happens for the lift coefficient, or hydrodynamic force coefficient, of a particle subject to a uniform flow with an orientation angle of~$\theta = 0^{o}$, $90^{o}$, or $180^{o}$.

The coefficient of determination is given by
\begin{equation}
\mathcal{R}^{2} = 1 - \frac{\sum_{i}^{N_\text{total}} \left( f_{\text{model},i} - f_{i}\right)^2 }{\sum_{i}^{N_\text{total}} \left(f_{i} - \bar{f}\right)}\, ,
\end{equation}
where $\bar{f}$ is the mean value of the distribution.
The closer the coefficient of determination is to $\mathcal{R}^{2} =1$, the better the model fit describes the variation in the set of data.

\subsection{Drag coefficient\label{sec:dragcoeff}}
To ensure a versatile correlation to predict the drag coefficient of a particle at varying distances to the wall, particularly when the particle is situated outside the wall boundary layer, PR-DNS results from~\citet{Cheron2024} for the case of an isolated fixed rod-like particle subjected to a uniform fluid flow are used in the limit for large values of dimensionless distances $\tilde{L}$.
When the PR-DNS results are not available for the desired flow case, the correlation to predict the drag coefficient in case of uniform flow is used~\citep{Cheron2024}.
Although the accuracy of the correlation is not assessed for particles with an aspect ratio lower than $\alpha < 2.5$, the correlations to predict the drag coefficient of a sphere in case of a wall-bounded linear shear flow of~\citet{Zeng2009} are used as a limit for the aspect ratio.

\subsubsection{The effect of the distance to the wall on the drag coefficient of a particle with a fixed orientation angle} 
To study the effect of the wall-distance $\tilde{L}$ on the drag coefficient of rod-like particles, insights from studies on spherical particles are briefly discussed.
\citet{Goldman1967a} demonstrate that for non-inertial flows the drag coefficient of a spherical particle fixed to a wall is proportional to the particle Reynolds number by a factor of
${40.81}{\left(\text{\Rep}\right)^{-1}}$, and decays as a function of the wall distance to the drag coefficient in case of Stokes flow, ${24}{\left(\text{\Rep}\right)^{-1}}$.\\

~\citet{Zeng2009} extend the work of~\citet{Goldman1967a} to consider finite Reynolds number effects, and show that for all particle Reynolds numbers a spherical particle fixed to a wall is always subject to a higher drag force compared to a particle subjected to a uniform flow.
In the work of~\citet{Zeng2009}, it is also shown that for all particle Reynolds numbers, \Rep, the drag coefficient of a particle  converges to the uniform flow drag coefficient as its distance from the wall increases.
Moreover, their results show that for intermediate distances and high particle Reynolds numbers, the drag coefficient of a spherical particle may be lower than the drag coefficient in case of uniform flow.
~\citet{Zeng2009} propose an empirical correlation to predict the drag coefficient of a spherical particle as a function of the distance to the wall, the particle Reynolds number, and the uniform flow drag coefficient,
\begin{equation}\label{eq:ZengDrag2009}
    \text{\cdl} = \left[\frac{24}{Re} \left(1 + 0.138 \exp\left(-2\tilde{L} + D_\text{eq}\right) + \frac{9}{16 ( 1 + 2\tilde{L} - D_\text{eq})} \right)\right]\left(1 + \beta_1 \text{\Rep}^{\beta_2}\right)\, ,\text{valid for }\alpha = 1\, ,
\end{equation}
where the first part of the right-hand side of the equation, enclosed in brackets, models the evolution of the drag coefficient as a function of the wall distance in the viscous regime, and is fitted from the results of~\citet{Goldman1967a}.
The second part of the equation, $\left(1 + \beta_1 \text{\Rep}^{\beta_2}\right)$, models the finite particle Reynolds number effects.
In Eq.~\eqref{eq:ZengDrag2009}, $\beta_1$ and $\beta_2$ are fit constants that vary as a function of $\tilde{L}$.
For very large distances to the wall,~Eq.~\eqref{eq:ZengDrag2009} simplifies to the drag correlation for spherical particles of~\citet{Schiller1933} in case of uniform fluid flow conditions.\\

\begin{figure}[ht!]
    \centering
    \includegraphics[width=0.995\columnwidth, trim={0 0 0 0}, clip]{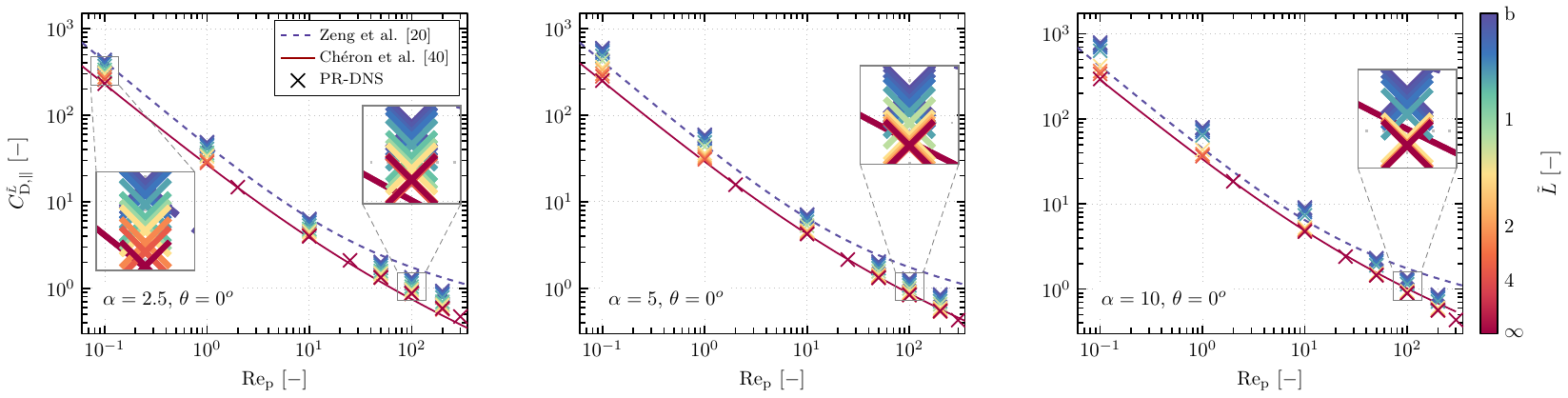}
    \caption{Evolution of the drag coefficient of particles at a fixed orientation angle of $\theta = 0^{o}$, as a function of the particle Reynolds number, \Rep, for different aspect ratios, $\alpha$, and distances to the wall, $\tilde{L}$. From left column to right column, $\alpha = 2.5$, $5$, and $10$. The colour map indicates the distance to the wall, from $\tilde{L} \rightarrow \infty$ (uniform flow case), to a particle fixed to the wall, $\tilde{L} = b$. \textcolor{black!50!red}{Solid red line}: drag correlation for rod-like particles at $\tilde{L} \rightarrow \infty$~\citep{Cheron2024}, \textcolor{Violet}{dashed blue line}: drag correlation for a sphere in a wall-bounded linear shear flow fixed to the wall, $\tilde{L} = b$~\citep{Zeng2009}.}
    \label{fig:dragcoefficient-evolution-Reynolds}
\end{figure}

To examine the variation of the drag coefficient for a non-spherical rod-like particle in case of wall-bounded linear shear flow compared to uniform flow, figure~\ref{fig:dragcoefficient-evolution-Reynolds} shows the evolution of the drag coefficient as a function of the particle Reynolds number, for several wall distances, for the three particle aspect ratios considered in this work. 
In this figure, the results for a fixed orientation angle, $\theta = 0^{o}$, are shown.
The drag coefficient is indicated with the symbol \cdpl, where the sub-script $\parallel$ indicates the fixed orientation angle of $\theta = 0^{o}$, and the super-script symbol $\tilde{L}$ indicates considering wall effects in the prediction of the drag coefficient. 
Thus, the absence of the super script $\tilde{L}$ indicates that the coefficient is predicted with correlations derived for uniform flow configurations.
Alongside the PR-DNS results of the present study, the PR-DNS results and drag correlation from~\citet{Cheron2024}, for similar particles subject to uniform fluid flow are shown, and represented with a wall distance of $\tilde{L} = \infty$. 
The correlation to predict the drag coefficient of a spherical particle of~\citet{Zeng2009}, see Eq.~\eqref{eq:ZengDrag2009}, is also included for a distance to the wall $\tilde{L} = b$, for a particle with an aspect ratio of $\alpha = 1$.
The results in figure~\ref{fig:dragcoefficient-evolution-Reynolds} show that, regardless of the aspect ratio or the particle Reynolds number, a closer proximity to the wall results in a larger drag coefficient. 
Conversely, when the particle is positioned further to the wall, the drag coefficient converges towards the values obtained for the case of uniform fluid flow. 
These trends are in good agreement with the findings of~\citet{Zeng2009} for spherical particles.
For particle Reynolds number \Rep = 0.1, the increase in the drag coefficient caused by the proximity of the particle to the wall is solely caused by the influence of the wall, since for this flow regime, the presence of an unbounded linear fluid velocity gradient does not increase the drag coefficient of a non-spherical particle~\citep{Harper1968,Cheron2024}\\

\begin{figure}[ht!]
    \centering
    \includegraphics[width=0.995\columnwidth, trim={0 0 0 0}, clip]{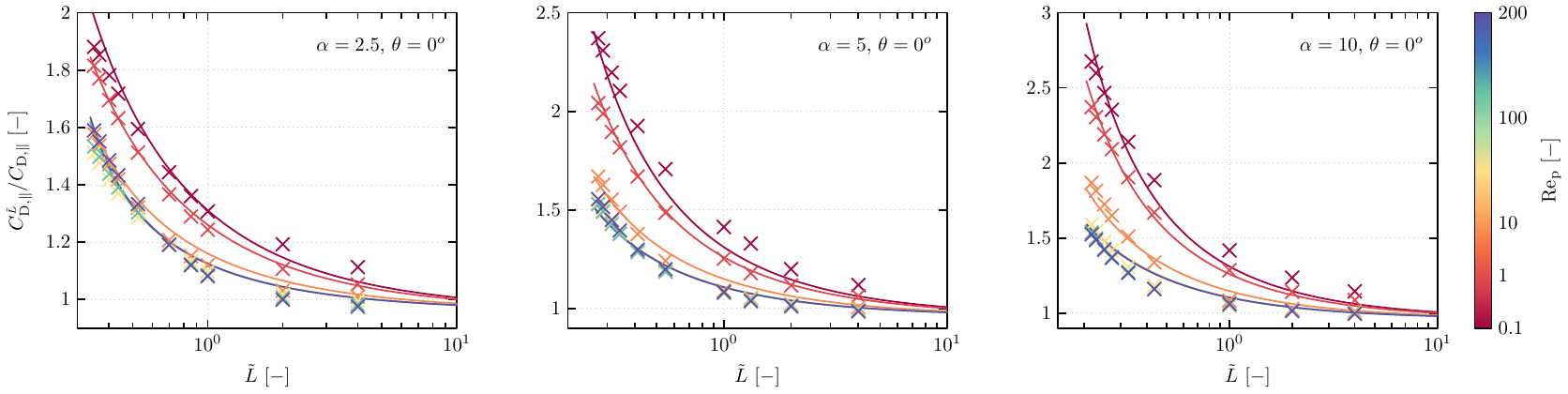}
    \caption{Evolution of the drag coefficient of a particle with a fixed orientation angle of $\theta = 0^{o}$, as a function of the distance to the wall, $\tilde{L}$, for different aspect ratios, $\alpha$, and particle Reynolds numbers, \Rep, \cdpl, scaled by the equivalent drag coefficient in case of uniform flow, \cdp. From left column to right column, $\alpha = 2.5$, $5$, and $10$. The colour map indicates the particle Reynolds number, \Rep. The markers : PR-DNS results, the solid lines: prediction given by Eq.~\ref{eq:drag-equation-distancetowall-first}.}
    \label{fig:dragcoefficient-evolution-Distance}
\end{figure}

The evolution of the drag coefficient as a function of the distance to the wall is shown in figure~\ref{fig:dragcoefficient-evolution-Distance} for several particle Reynolds numbers. The drag coefficient is scaled by the drag coefficient in case of uniform flow ($\tilde{L} \rightarrow \infty$). 
The results are shown for an orientation angle of $\theta = 0^{o}$, and for the particles with aspect ratios of $\alpha = 2.5, 5$ and $10$.
For all particles and particle Reynolds numbers, the closer the particle is to the wall, the larger is the change in the drag coefficient compared to uniform flow.
Moreover, the magnitude of the change in the drag coefficient is inversely related to the particle Reynolds number; lower Reynolds numbers lead to larger changes in the drag coefficient.
The aspect ratio of the non-spherical particle also significantly influences the magnitude of the change in the drag coefficient.
For instance, at \Rep = 0.1 and $\tilde{L} = b$,~\citet{Goldman1967a} report that the increase in the drag coefficient for a spherical particle is a factor 1.7.
In the present configuration, the increase of the drag coefficient for a non-spherical particle fixed to the wall at an orientation angle of $\theta = 0^{o}$ is of a factor 1.88, 2.37, and 2.68, for the aspect ratios $\alpha = 2.5$, 5 and 10, respectively.
At higher particle Reynolds number, for instance \Rep $\geq 10$, the evolution of the drag coefficient as a function of the wall distance, scaled by the drag coefficient in case of uniform flow, overlaps for the particles with aspect ratios of $\alpha = 2.5$ and 5. This overlap is also observed for spherical particles~\citep{Zeng2009}.
For particle Reynolds numbers exceeding \Rep $\geq 10$, the drag coefficient of a particle fixed to the wall at an orientation angle of $\theta = 0^{o}$ increases by approximately a factor of 1.5 compared to the drag coefficient in uniform flow. In the case of a particle with an aspect ratio of $\alpha = 10$, the influence of the particle Reynolds number on the change in the drag coefficient diminishes for particle Reynolds numbers equal to or greater than \Rep $\geq 50$ when compared to the drag coefficient in uniform flow.
For a particle with aspect ratio $\alpha = 10$, the particle Reynolds number no longer influences the magnitude of the change in the drag coefficient when compared to the drag coefficient in case of uniform flow for values of particle Reynolds numbers equal or greater than \Rep $\geq 50$.
Thus, as the particle becomes more spherical, the influence of the particle Reynolds number on the maximum value of the drag coefficient of a particle fixed to a wall, scaled by the uniform flow drag coefficient, decreases.\\

This observation suggests the possibility of a general form for the fitting model predicting the evolution of the drag coefficient of a particle with orientation angle $\theta = 0^{o}$, as a function of the wall distance, valid for all particle Reynolds numbers and aspect ratios considered in this work.
This correlation is derived based on the uniform flow correlation for the drag coefficient of~\citet{Cheron2024}.
As shown in figure~\ref{fig:dragcoefficient-evolution-Distance}, this model needs to capture the specific dependence between the influence of the aspect ratio and the particle Reynolds number, when the latter decreases. 
The proposed correlation to predict the drag coefficient of a particle with a fixed orientation angle, of $\theta = 0^{o}$, which varies as a function of the aspect ratio, $\alpha$, the distance to the wall, $\tilde{L}$, and the particle Reynolds number, \Rep, is given by
\begin{equation}\label{eq:drag-equation-distancetowall-first}
\text{\cdpl}= \text{\cdp} \left[
    \exp\left(\mathcal{D}_1(\alpha,\tilde{L},\text{\Rep}) \mathcal{D}_2\left(\text{\Rep}\right)^{-1}\right) +\mathscr{\delta}_1 \right]\, ,
\end{equation}
where the functions $\mathcal{D}_1$ and $\mathcal{D}_2$ express the specific profile observed in figure~\ref{fig:dragcoefficient-evolution-Distance}. These functions are given by
\begin{align}\label{eq:drag-equation-distancetowall}
&\mathcal{D}_1\left(\alpha,\tilde{L},\text{\Rep}\right) = \mathscr{\delta}_2 \left[\left(\tilde{L} - (\mathscr{\delta}_3^{ \left(\mathscr{\delta} \right) })^{\left(\mathscr{\delta}_6\right)}\right)\right]^{-1}\, ,\quad\text{with}\quad \mathscr{\delta} = \left[\frac{1}{\text{\Rep} \mathscr{\delta}_4} + \alpha\right] - \mathscr{\delta}_5\, ,\\
&\mathcal{D}_2\left(\text{\Rep}\right) = \mathscr{\delta}_7 - \mathscr{\delta}_8^{\text{\Rep}^{\mathscr{\delta}_9}}
\end{align}

The fit parameters in Eq.~\eqref{eq:drag-equation-distancetowall} to predict the change in the drag coefficient at a fixed orientation angle of $\theta = 0^{o}$, compared to the drag coefficient in case of uniform flow, have been determined with a fitting algorithm and are listed in Table~\ref{table:dragcoefficient-zero}. Although an underestimation of the prediction of the drag coefficient at low particle Reynolds number is observed, the correlations accurately recover the evolution of the drag coefficient of the non-spherical particle over the entire range of particle Reynolds numbers, distances to the wall, and aspect ratios.

\subsubsection{The effect of the orientation of the particle on the drag coefficient}

The derivation of a correlation to predict the drag coefficient of an axi-symmetric particle subject to locally uniform fluid flow often relies on the ``sinesquare'' profile, which predicts the evolution of the drag coefficient as a function of the orientation angle based on the drag coefficient at orientation angles $\theta = 0^{o}$ and $90^{o}$, \cdp{} and \cdpp, respectively.
This profile is initially formulated for non-inertial flows by~\citet{Brenner1963}, and is given by
\begin{equation} 
    \text{\cd} = \text{\cdp} + \left[\text{\cdpp} - \text{\cdp}\right]\sin(\theta)^{2}\, .
\end{equation}
The ``sinesquare'' profile has been demonstrated to be a satisfactory assumption for predicting the drag coefficient of axi-symmetric prolate, and rod-like particles subjected to a locally uniform flow at high particle Reynolds numbers~\citep{Sanjeevi2018, Cheron2024}.
However, more complex fitting correlations are also suggested for rod-like particles~\citep{Feng2023}.\\

\begin{figure}[ht!]
    \centering
    \begin{tabular}{m{0.03\linewidth}m{0.95\linewidth}}
    \rotatebox{90}{$\quad$ \Rep = 0.1} &
     \includegraphics[width=0.95\columnwidth, trim={0 0 0 0}, clip]{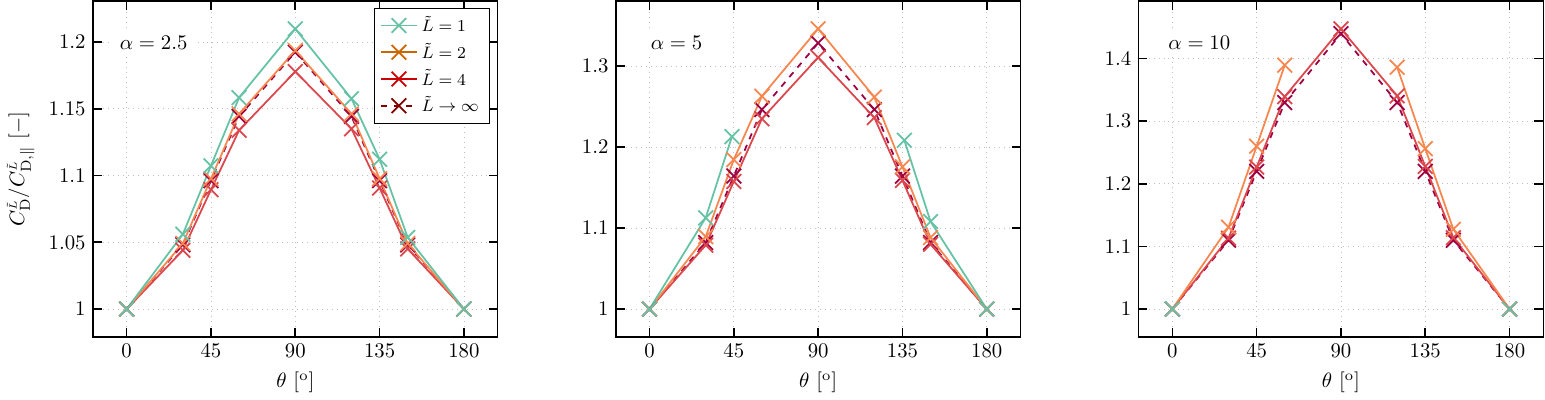}\\
    \rotatebox{90}{$\quad$ \Rep = 10} &
     \includegraphics[width=0.95\columnwidth, trim={0 0 0 0}, clip]{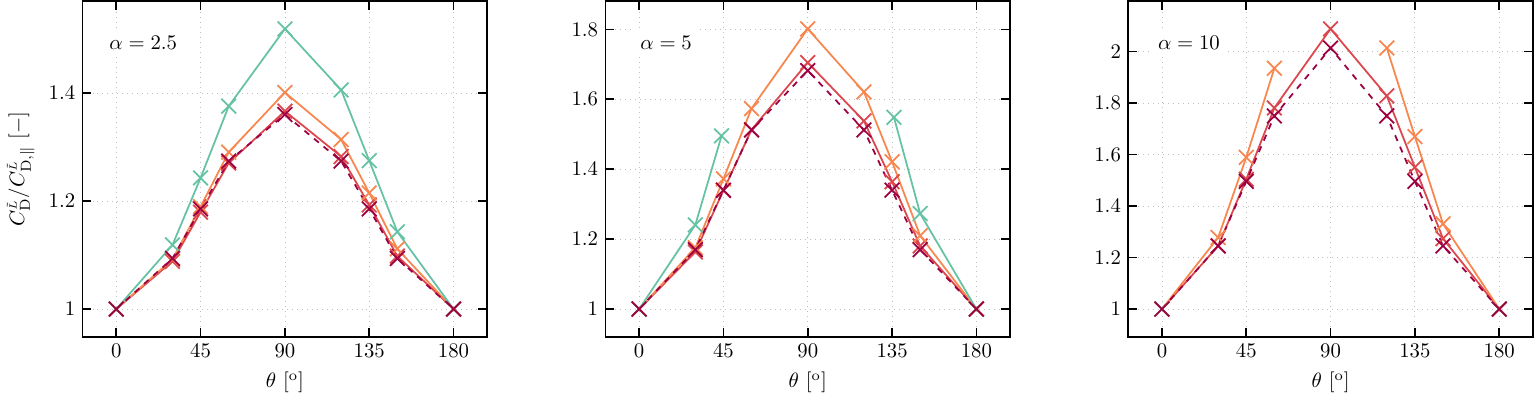}\\
    \rotatebox{90}{$\quad$ \Rep = 200} &
     \includegraphics[width=0.95\columnwidth, trim={0 0 0 0}, clip]{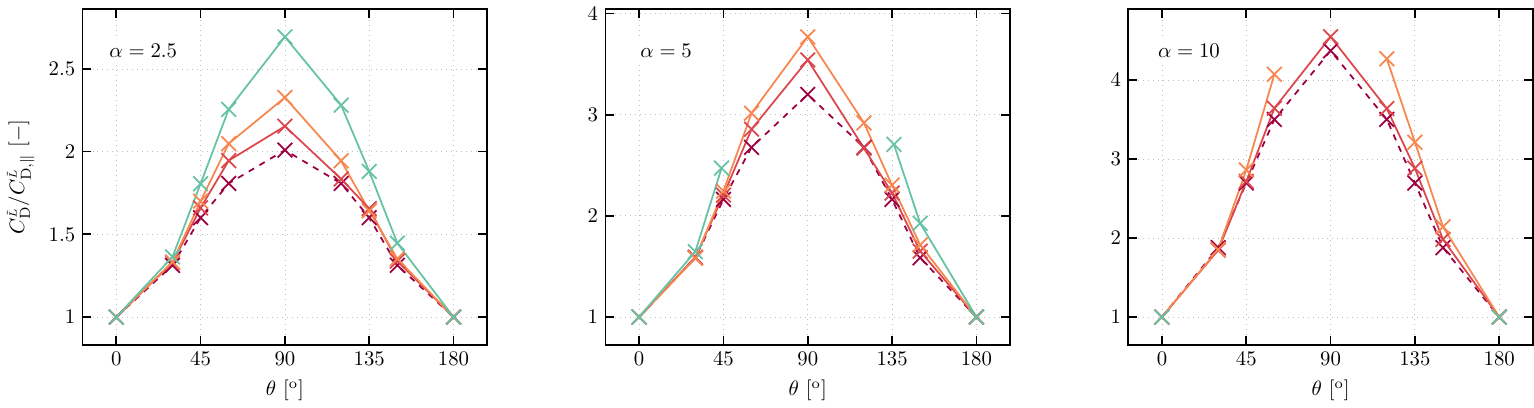}\\
    \end{tabular}
    \caption{Evolution of the change in the drag coefficient of a particle due to the proximity to the wall, $C_\textup{D}$, scaled by the equivalent drag coefficient at a fixed orientation angle of $\theta = 0^{o}$, $C_\textup{D}\left(\theta = 0^{o}\right)$, as a function of the orientation angle, $\theta$, for different aspect ratios, $\alpha$, particle Reynolds numbers, \Rep, and distances to the wall, $\tilde{L}$. From left column to right column, $\alpha = 2.5$, $5$, and $10$. From top row to bottom row, \Rep = 0.1, 10 and 200. Colour of solid lines  with markers indicates distance to the wall, $\tilde{L} = 4$ : \textcolor{black!20!red}{red solid line}, $\tilde{L} = 2$ : \textcolor{black!20!orange}{orange solid line}, $\tilde{L} = 1$ : \textcolor{blue!40!green}{green solid line}. Dashed line indicates drag coefficient in case of uniform flow~\citep{Cheron2024}.}
    \label{fig:dragcoefficient-evolution}
\end{figure}

In the context of this study, the value of the drag coefficient at an orientation angle of $\theta = 90^{o}$ is not accessible for all dimensionless distances from the wall.
Thus, the ``sinesquare'' expression is not a good candidate to model the change in the drag coefficient in case of wall-bounded linear shear flow, compared to uniform flow, as a function of the orientation of the particle.
However, the comparison with the drag coefficient in case of uniform flow can be done by scaling the drag coefficient in case of wall-bounded linear shear flow by the drag coefficient for the equivalent simulation case with an orientation angle of $\theta = 0^{o}$.
This is shown in figure~\ref{fig:dragcoefficient-evolution}, for the particle Reynolds numbers \Rep = 0.1, 10 and 200, the dimensionless distances to the wall $\tilde{L} = \infty$, 4, 2 and 1, and all the aspect ratios considered in this study.\\

At low particle Reynolds number, \Rep = 0.1, the profile of the scaled drag coefficient evolution as a function of the orientation angle closely resembles that of the drag coefficient obtained in a uniform flow.
Notably, the most significant increase in the drag coefficient occurs at an orientation angle of $\theta = 90^{o}$.
Moreover, the magnitude of this deviation increases with the decrease in the distance between the particle and the wall.
The aspect ratio also plays a role, the larger is the aspect ratio, the larger is the change at $\theta = 90^{o}$.
In addition, the profile of the scaled drag coefficient evolution remains symmetric about the orientation angle $\theta = 90^{o}$.
Thus, at low particle Reynolds number, \Rep = 0.1, modelling the drag coefficient evolution in the range $\theta = 0^{o}$ to $90^{o}$ adequately represents all orientation angles examined in this study.\\

However, at higher particle Reynolds numbers, \Rep = 10 and 200, the evolution of the drag coefficient as a function of the orientation angle scaled by the drag coefficient in case of uniform flow no longer maintains its symmetry at the orientation angle $\theta = 90^{o}$. For example, at \Rep = 10, for all aspect ratios, the drag coefficient consistently exceeds that for orientation angles between $\theta = 90^{o}$ to $180^{o}$ compared to $\theta = 0^{o}$ to $90^{o}$. This specific profile is observed in the results for a particle with an aspect ratio of $\alpha = 5$ and dimensionless distances of $\tilde{L} = 1$ and $2$. At higher particle Reynolds numbers, \Rep = 200, the trend becomes more complex, and varies with the distance to the wall and the aspect ratio of the particle.
This distinctive pattern is also observed for particles subjected to an unbounded linear shear flow~\citep{Cheron2024}.\\

\begin{figure}[ht!]
    \centering
    \begin{tabular}{m{0.03\linewidth}m{0.95\linewidth}}
    \rotatebox{90}{$\quad$ \Rep = 0.1} &
    \includegraphics[width=0.84\columnwidth, trim={50 400 50 50}, clip]{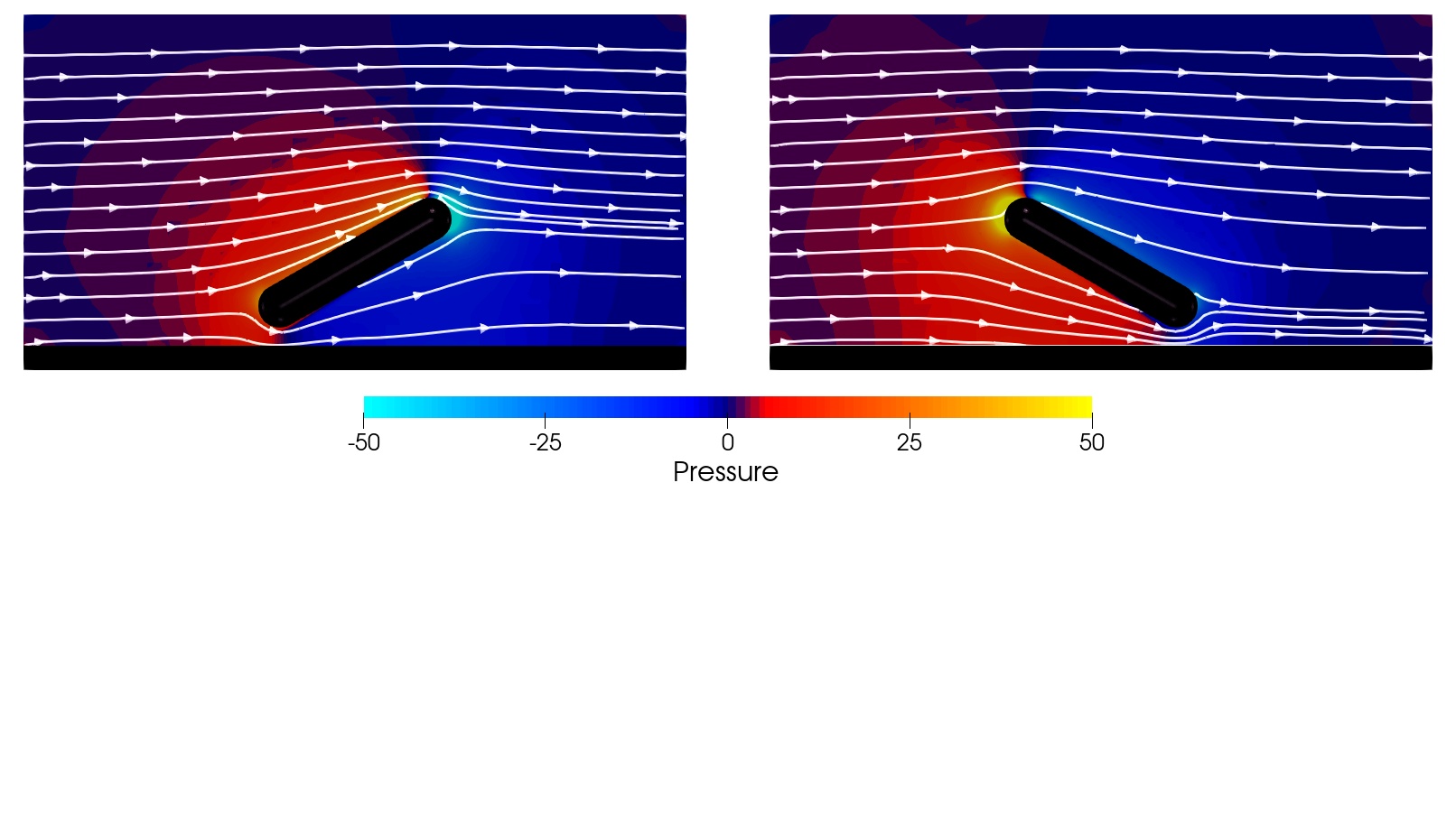}\\[2mm]
    \rotatebox{90}{$\quad$ \Rep = 100} &
    \includegraphics[width=0.84\columnwidth, trim={50 400 50 100}, clip]{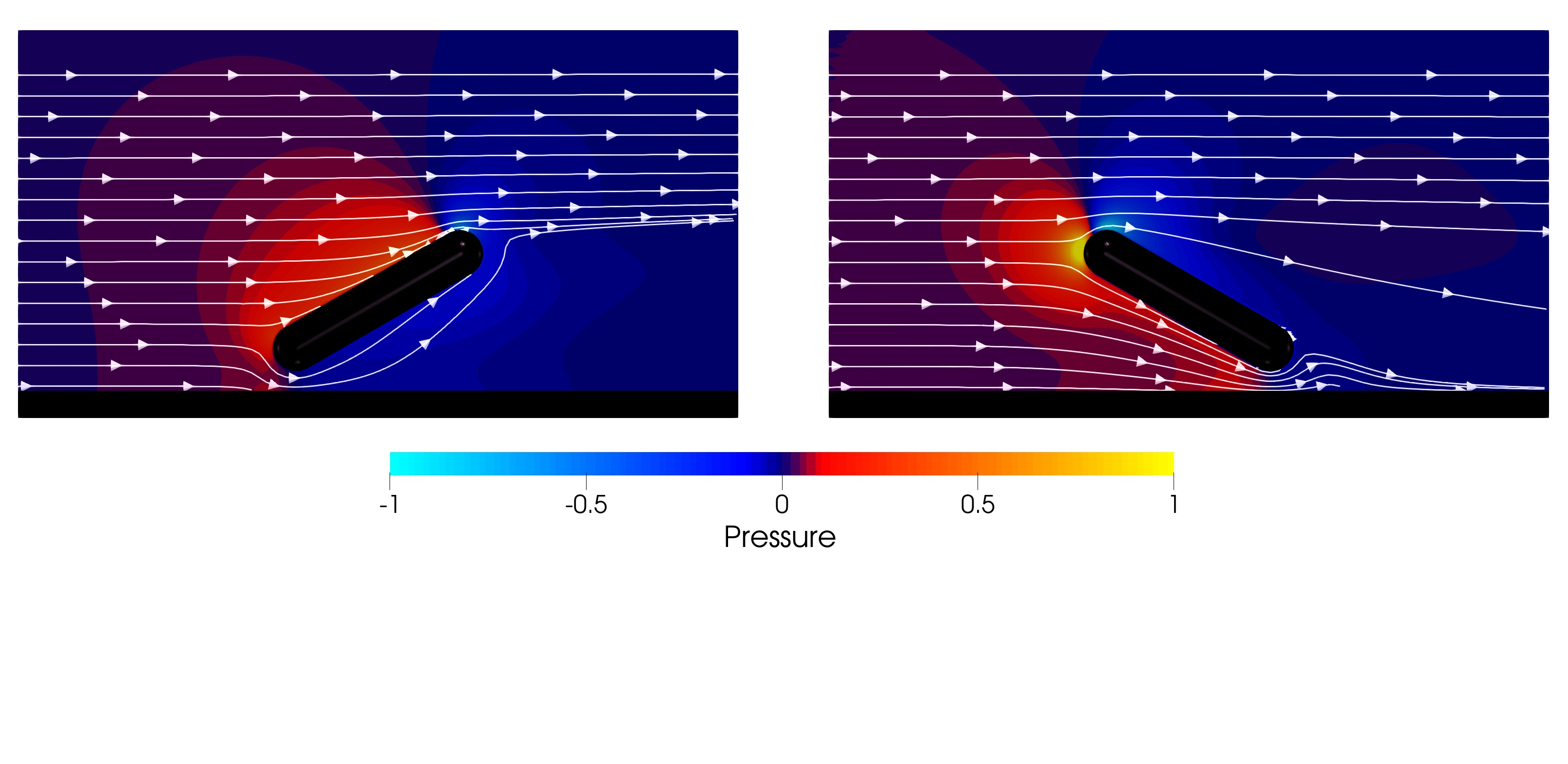}
    \end{tabular}
    \caption{Snapshots of the magnitude of the fluid velocity (streamlines), with pressure contours (colour map), past a fixed rod-like particle of aspect ratio $\alpha = 5$, at a dimensionless distance from the wall $\tilde{L} = 1$. Top row particle Reynolds number \Rep = 0.1, bottom row \Rep = 100. Left column, orientation angle $\theta = 30^{o}$, right column $\theta = 150^{o}$.}
    \label{fig:flow-evolution}
\end{figure}

The loss of symmetry in the evolution of the drag coefficient as a function of the orientation angle at the orientation angle $\theta = 90^{o}$ can also be observed through qualitative analysis of the PR-DNS results. In figure~\ref{fig:flow-evolution}, the instantaneous fluid velocity and pressure fields are shown for a particle with aspect ratio $\alpha = 5$ at a dimensionless distance from the wall $\tilde{L} = 1$, for orientation angles $\theta = 30^{o}$ and $150^{o}$ (left and right columns), and particle Reynolds numbers \Rep = 0.1 and 100 (top and bottom rows).
At a particle Reynolds number of \Rep = 0.1, the pressure field distribution near the surface of the particle for orientation angles $\theta = 30^{o}$ and $150^{o}$ exhibit significant similarity. For instance, the stagnation point is located at the top of both particles, and the maximum drop in pressure is equivalent.
However, at \Rep = 100, the pressure distribution varies with the orientation of the particle. For instance, at an orientation angle of $\theta = 30^{o}$, the pressure distribution resembles that observed at \Rep = 0.1. However, this is not the case for an orientation angle of $\theta = 150^{o}$; the absence of a low-pressure point in the wake of the particle leads to the loss of symmetry in the evolution of the drag coefficient with respect to orientation angle.\\

In addition to the loss of symmetry, an increase in the particle Reynolds number results in an increase in the change in the drag coefficient of the particle near the wall, compared to the uniform flow configuration.
For example, for a particle with an aspect ratio of $\alpha = 2.5$, at a particle Reynolds number of \Rep = 200, a dimensionless distance to the wall of $\tilde{L} = 1$, and an orientation angle of $\theta = 90^{o}$, the drag coefficient in case of wall-bounded linear shear flow, scaled by the drag coefficient for the equivalent simulation case with an orientation angle of $\theta = 0^{o}$, is 1.37 times larger than in the case of uniform flow, whereas the difference is negligible for the equivalent configuration at a particle Reynolds number of \Rep = 0.1.
The evolution of the scaled drag coefficient shown in figure~\ref{fig:dragcoefficient-evolution} suggests initially modelling the change in the drag coefficient as a function of the uniform flow coefficient, and the drag coefficient in case of wall-bounded linear shear flow at a fixed orientation angle of $\theta = 0^{o}$, for low particle Reynolds numbers, \Rep = 0.1. Finite Reynolds number effects can then be incorporated into the expression to accurately predict the change in the drag coefficient at high particle Reynolds numbers.\\

\begin{figure}[ht!]
    \centering
    \begin{tabular}{m{0.03\linewidth}m{0.95\linewidth}}
    \rotatebox{90}{$\quad$ \Rep = 0.1} &
     \includegraphics[width=0.95\columnwidth, trim={0 0 0 0}, clip]{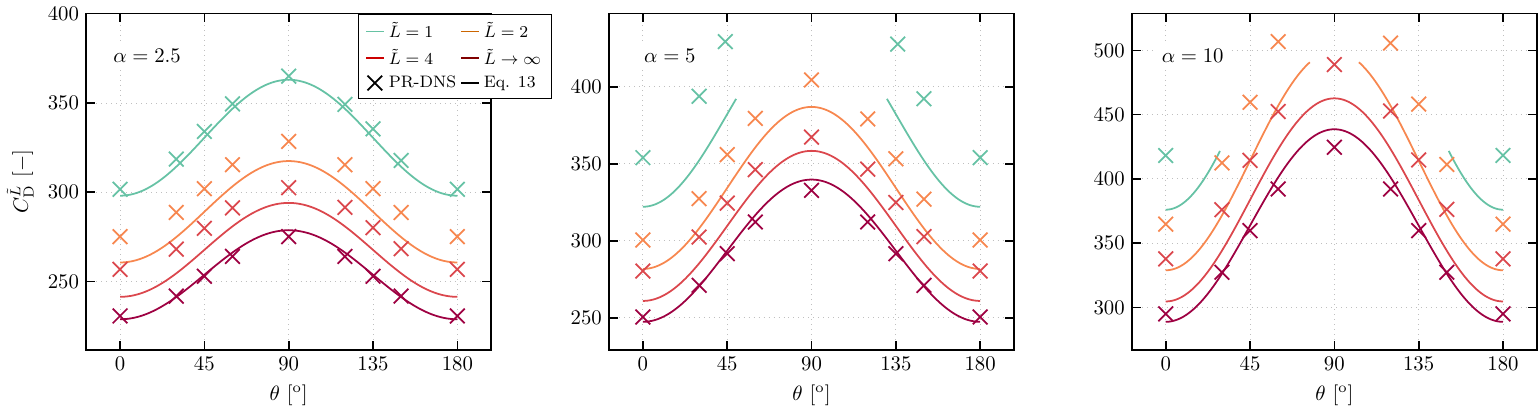}\\
    \rotatebox{90}{$\quad$ \Rep = 10} &
     \includegraphics[width=0.95\columnwidth, trim={0 0 0 0}, clip]{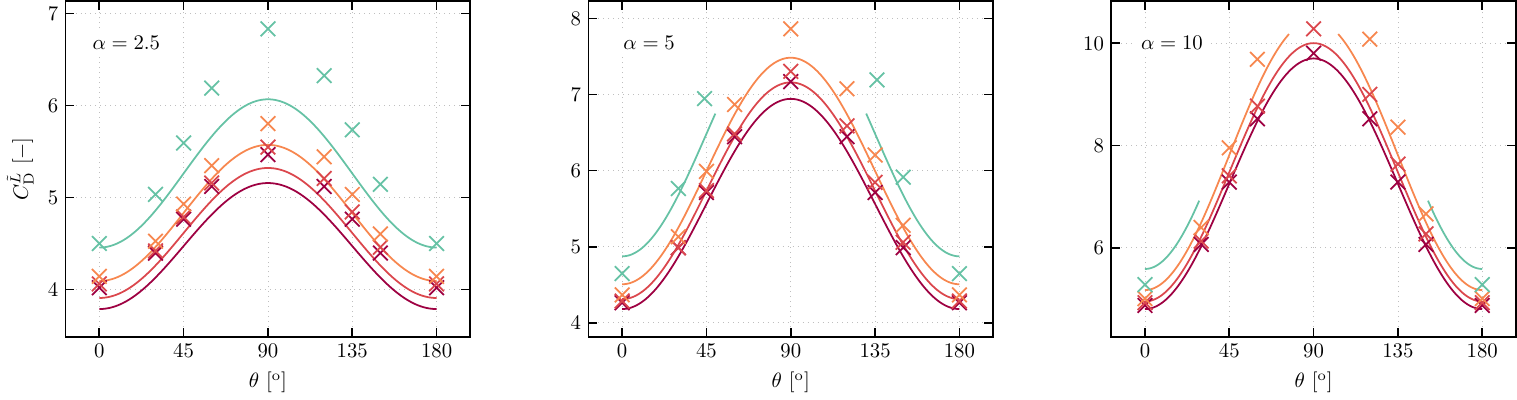}\\
    \rotatebox{90}{$\quad$ \Rep = 200} &
     \includegraphics[width=0.95\columnwidth, trim={0 0 0 0}, clip]{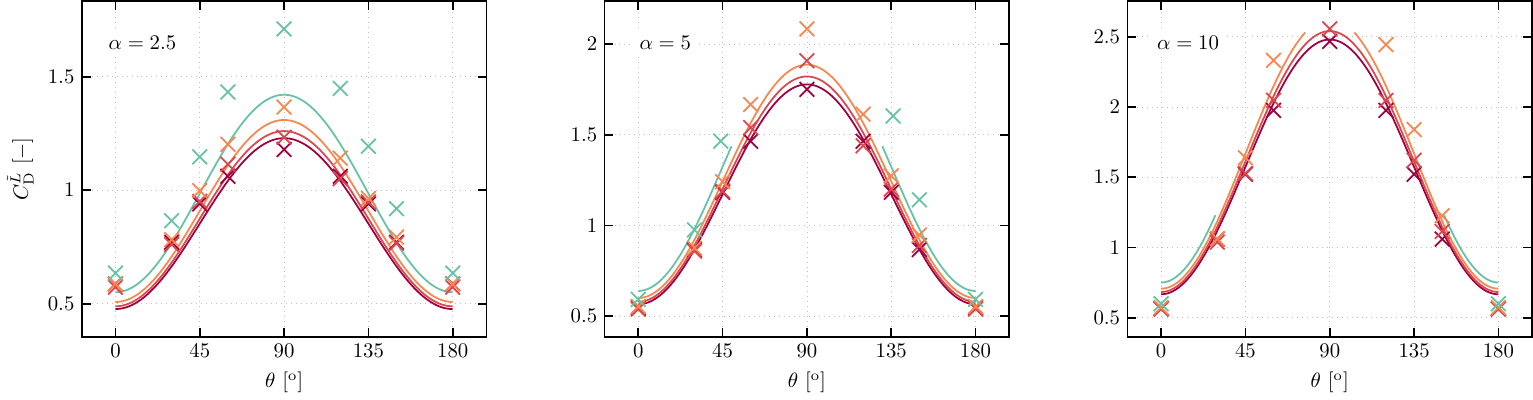}\\
    \end{tabular}
     \caption{Evolution of the drag coefficient, \cd, as a function of the orientation angle, $\theta$. From left column to right column, $\alpha = 2.5$, $5$, and $10$.  From top row to bottom row, \Rep = 0.1, 10 and 200. Markers: PR-DNS results, solid line: prediction given by Eq.~\ref{eq:drag-equation-distancetowall-all}. Colour indicates distance to the wall, $\tilde{L} = \infty$ : \textcolor{black!50!red}{dark red}, $\tilde{L} = 4$ : \textcolor{black!20!red}{red}, $\tilde{L} = 2$ : \textcolor{black!20!orange}{orange}, $\tilde{L} = 1$ : \textcolor{blue!40!green}{green}.}
    \label{fig:dragcoefficient-fit}
\end{figure}

To analyse the change in the drag coefficient compared to the drag coefficient in case of uniform flow, figure~\ref{fig:dragcoefficient-fit} shows the evolution of the drag coefficient in case of wall-bounded linear shear flow as a function of the orientation angle for the results shown in figure~\ref{fig:dragcoefficient-evolution}.
For a particle Reynolds number of \Rep = 0.1, all the aspect ratios, and a dimensionless distance of $\tilde{L} = 4$, the evolution of the scaled drag coefficient as a function of the orientation angle remains almost constant.
This suggests that the correlation proposed to model the drag coefficient at a fixed orientation angle of $\theta = 0^{o}$, given in Eq.~\eqref{eq:drag-equation-distancetowall-first}, can effectively predict the drag coefficient for all orientation angles, since the orientation of the particle does not magnify the change in the coefficient compared to the change observed with an orientation angle of $\theta = 0^{o}$.
For shorter distances, such as $\tilde{L} = 1$ or $\tilde{L} = 2$. and for the particle with a larger aspect ratio, the evolution of the change in the drag coefficient is no longer constant for all orientation angles. 
The largest change with respect to the case with uniform flow is observed for the orientation angle of $\theta = 90^{o}$, and this change increases linearly from $\theta = 0^{o}$ to $90^{o}$, with a similar symmetric evolution in the range $\theta = 180^{o}$ to $90^{o}$.\\

With an increase in the particle Reynolds number, see for instance \Rep = 10 in figure~\ref{fig:dragcoefficient-fit}, the magnitude of the change in the drag coefficient in case of linear-wall bounded flow compared to the case with uniform flow decreases for all orientation angles.
Also, the evolution of the drag coefficient as a function of the orientation angle is no longer symmetric at $\theta = 90^{o}$, the change in the drag coefficient in case of linear-wall bounded flow is more prominent for particles with an orientation angle ranging from $\theta = 180^{o}$ to $90^{o}$ than $\theta = 0^{o}$ to $90^{o}$.
This effect is amplified for particles with larger aspect ratios, and shorter distances to the wall, as shown qualitatively for the fluid flow fields in figure~\ref{fig:flow-evolution}.
For larger particle Reynolds number, \Rep = 200, for particle with aspect ratios $\alpha = 2.5$ and $5$, the intricate evolution of the fluid flow around the particle situated at a few dimensionless distances away from the wall, in this study at $\tilde{L} = 2$ and $4$, leads to a more prominent increase in the drag coefficient for a particle with orientation angles in the range $\theta = 0^{o}$ to $90^{o}$ than $\theta = 180^{o}$ to $90^{o}$.
However, this is not observed for the particle with the largest aspect ratio, $\alpha = 10$, and for the shortest distances to the wall, $\tilde{L} \leq 1$.\\

To capture the influence of the orientation of the particle with respect to the local fluid flow direction, a first correlation is given as a function of the prediction of the drag coefficient of a particle parallel to the wall for low particle Reynolds number.
Finite Reynolds number effects are incorporated based on the expression derived for low particle Reynolds number, considering the distinct profile of the evolution of the drag coefficient as a function of the orientation angle in the ranges $\theta = 0^{o}$ to $90^{o}$ and $\theta = 90^{o}$ to $180^{o}$.
This correlation is given by
\begin{equation}\label{eq:drag-equation-distancetowall-all}
\text{\cdl} = \text{\cdpl}\left[1 + \left(\cfrac{\mathscr{\delta}_1 \alpha^{\mathscr{\delta}_2}\sin\left(\Psi\left(\theta,\alpha,\tilde{L}\right)\right)^2}{\mathscr{\delta}_3 \tilde{L}^{\mathscr{d\delta}}}\right)\gamma\left(\text{\Rep},\tilde{L}\right)\right]\, ,\quad\text{with}\quad \mathscr{\delta} = \mathscr{\delta}_4 + \mathscr{\delta}_5 \alpha\, ,
\end{equation}
where \cdpl is given in Eq.~\eqref{eq:drag-equation-distancetowall-first}, and $\gamma$ and $\Psi$ are functions which express the change in the drag coefficient caused by the finite particle Reynolds number effects, and are given by
\begin{equation}
\gamma\left(\text{\Rep},\tilde{L}\right) = \mathscr{\delta}_6 \exp\left(\mathscr{\delta}_{7}\text{\Rep}^{\left(\mathscr{\delta}_{8}\tilde{L}\right)}\right)\, ,
\end{equation}
and the function $\Psi$ is derived to model the loss of symmetry of the drag coefficient at the orientation angle $\theta = 90^{o}$, and is given by
\begin{align}\label{eq:CD-wallshear-fit-Rep}
\Psi\left(\theta,\alpha,\tilde{L}\right) = \mathscr{d}_{\Psi} = \left\{
    \begin{array}{ll}
    \theta & \text{\Rep} < 1\, ,\\
     \frac{\pi}{2}\left(\frac{\theta}{\pi/2}\right)^{1 + \mathscr{\delta}_9 \tilde{L}} & \text{\Rep} \geq 1 \quad\text{and}\quad \theta < \nicefrac{\pi}{2}\, ,\\
     \frac{-\pi}{2}\left(\frac{\pi - \theta}{\pi/2}\right)^{1 + (\mathscr{\delta}_{10} \alpha \tilde{L})^{\mathscr{\delta}_{11} } } & \text{\Rep} \geq 1 \quad\text{and}\quad \theta \geq \nicefrac{\pi}{2}\, .
    \end{array}
\right.
\end{align}
The fit parameters for the Eq.~\eqref{eq:drag-equation-distancetowall-all} are listed in table~\ref{table:dragcoefficient-zero} under the appropriate equation number.

While some predictions of the drag coefficient underestimate the values obtained with the PR-DNSs, particularly for results at high particle Reynolds numbers or for some cases such as \Rep = 0.1, $\alpha = 2.5$, and $\tilde{L} = 2$ and $4$, overall, the model to predict the drag coefficient shows a good agreement over the whole range of parameters.
For instance, the median relative error between the model prediction and the PR-DNS results of ${\mathcal{E}} = 2.89\%$ is obtained.
In addition, the correlation coefficient between the model prediction and the results is equal to $\mathcal{R}^{2} = 0.99$, showing a very good agreement for the whole data set.

\begin{table}
\centering
\resizebox{\columnwidth}{!}{%
\begin{tabular}{l | c c c c c c c c c c c}
 & $\mathscr{\delta}_1$ & $\mathscr{\delta}_2$ & $\mathscr{\delta}_3$ & $\mathscr{\delta}_4$ & $\mathscr{\delta}_5$ & $\mathscr{\delta}_6$ & $\mathscr{\delta}_7$ & $\mathscr{\delta}_8$ & $\mathscr{\delta}_9$ & $\mathscr{\delta}_{10}$ & $\mathscr{\delta}_{11}$ \\
\hline
\hline\\[-1mm]
Eq.~\eqref{eq:drag-equation-distancetowall-first} & $-4.09 \times 10^{-2}$ & 0.239 & 0.529 & $8.990 \times 10^{-2}$ & -0.456 & 0.811 & 1.769 & 0.862 & 0.811 &  - & -\\
Eq.~\eqref{eq:drag-equation-distancetowall-all} & $-1.400 \times 10^{-3}$ & 0.8656& -4.4844 &  2.4368 &  -0.1033 &  257.1673 & -6.4265 & -5.5923 & 0.5774& 0.3607& -3.1566\\
\end{tabular}
}
     \caption{List of the fit parameters in Eqs.~\eqref{eq:drag-equation-distancetowall-first} and ~\eqref{eq:drag-equation-distancetowall-all}, used in the correlation to predict the change in the drag coefficient in case of wall-bounded linear shear flow, with respect to the uniform flow drag coefficient.}\label{table:dragcoefficient-zero}
\end{table}

\subsubsection{Summary of drag coefficient correlation and range of validity\label{sec:drag-correl}}

The correlations derived to predict the change in the drag coefficient in case of wall-bounded linear shear flow, compared to the drag coefficient in case of uniform flow, are summarized in the table~\ref{table:dragcoefficient-summary}.
This correlation varies as a function of the particle Reynolds number, \Rep, the orientation angle between the particle major axis and the local direction of the fluid flow, $\theta$, the aspect ratio of the particle, $\alpha$, and the dimensionless distance between the centre of the particle and the wall $\tilde{L}$.
The correlation to predict the change in the drag coefficient of a non-spherical particle in a wall-bounded linear shear flow is valid for particle Reynolds number in the range 0.1 $\leq$ \Rep  $\leq$ 200, orientation angle in the range  $0^{o}\leq \theta \leq 180^{o}$, where the orientation angle varies in the fluid flow direction and wall-normal direction plane, particle with an aspect ratio in the range $2.5 \leq \alpha \leq 10$, and dimensionless distances from touching the wall, $\tilde{L} = b$, to outside the boundary layer $\tilde{L} \rightarrow \infty$. The latter condition ensures that, as $\tilde{L}$ increases, the present correlation asymptotically approaches the correlation predicting the drag coefficient of the specific particle in case of locally uniform flow~\citep{Cheron2024}, as suggested by~\citet{Zeng2009} that put forward a correlation predicting the drag coefficient of a spherical particle under similar flow conditions.

\begin{table}
  \centering
  \begin{tabular}{l | r c l l}
     & Coefficient &  & Formula & \\
    \hline
    \hline
    {Eq.~\eqref{eq:drag-equation-distancetowall-all}} & {\cdl/\cd} & {=} & {$\text{\cdpl}\left[1 + \left(\cfrac{\mathscr{\delta}_1 \alpha^{\mathscr{\delta}_2}\sin\left(\Psi\left(\theta,\alpha,\tilde{L}\right)\right)^2}{\mathscr{\delta}_3 \tilde{L}^{\mathscr{d\delta}}}\right)\gamma\left(\text{\Rep},\tilde{L}\right)\right] $} & {$\text{with}\quad \mathscr{\delta} = \mathscr{\delta}_4 + \mathscr{\delta}_5 \alpha$}\\[4mm]
     - & $\gamma\left(\text{\Rep},\tilde{L}\right) $ & =  & $\mathscr{\delta}_6 \exp\left(\mathscr{\delta}_{7}\text{\Rep}^{\left(\mathscr{\delta}_{8}\tilde{L}\right)}\right)$\\[4mm]
     - & $\Psi\left(\theta,\alpha,\tilde{L}\right)$ & = & $\theta$ & for \Rep $\leq 1$ \\[4mm]
    - & $\Psi\left(\theta,\alpha,\tilde{L}\right)$ & =  & $\frac{\pi}{2}\left(\frac{\theta}{\pi/2}\right)^{1 + \mathscr{\delta}_9 \tilde{L}}$ & for \Rep $> 1$ and $\theta < \nicefrac{\pi}{2}$\\[4mm]
    - & $\Psi\left(\theta,\alpha,\tilde{L}\right)$ & =  & $\frac{-\pi}{2}\left(\frac{\pi - \theta}{\pi/2}\right)^{1 + (\mathscr{\delta}_{10} \alpha \tilde{L})^{\mathscr{\delta}_{11} } }$ & for \Rep $> 1$ and $\theta \geq \nicefrac{\pi}{2}$\\[4mm]
    \hline\\
    {~\citep[Eq. 27]{Cheron2024}} & {\cd}  & {}& {\citet{Cheron2024}} & {}\\[1mm]
    \hline\\
    {Eq.~\eqref{eq:drag-equation-distancetowall-first}} & {\cdpl/\cdp} & {=} & {$ \left[
      \exp\left(\mathcal{D}_1(\alpha,\tilde{L},\text{\Rep}) \mathcal{D}_2\left(\text{\Rep}\right)^{-1}\right) +\mathscr{\delta}_1 \right]$} & {$\text{with}\quad \mathscr{\delta} = \mathscr{\delta}_4 + \mathscr{\delta}_5 \alpha$}\\[4mm]
      - & $\mathcal{D}_1$ & = & $\mathscr{\delta}_2 \left[\left(\tilde{L} - (\mathscr{\delta}_3^{ \left(\mathscr{\delta} \right) })^{\left(\mathscr{\delta}_6\right)}\right)\right]^{-1}$ & $\mathscr{\delta} = \left[\frac{1}{\text{\Rep} \mathscr{\delta}_4} + \alpha\right] - \mathscr{\delta}_5 $\\[4mm]
      - & $\mathcal{D}_2$ & = & $\mathscr{\delta}_7 - \mathscr{\delta}_8^{\text{\Rep}^{\mathscr{\delta}_9}}$ &  \\[4mm]
      \hline\\
      {~\citep[Eq. 27]{Cheron2024}} & {\cd}  & {}& {\citet{Cheron2024}} & {}\\[1mm]
  \end{tabular}
       \caption{Correlation to predict the drag coefficient of a particle subjected to a wall-bounded linear shear flow as a function of the particle Reynolds number, \Rep, the orientation angle between the major axis of the particle and the local fluid flow direction, $\theta$, the aspect ratio of the particle, $\alpha$, and the dimensionless distance between the centre of the particle and the wall, $\tilde{L}$. The drag coefficient, {\cdl}, is given as a function of the drag coefficient for a particle subject to uniform flow condition, \cd, and the coefficients of the same particle with an orientation angle fixed to $\theta = 0^{o}$ for both flow conditions, {\cdpl}, and {\cdp}. The fit parameters, $\mathscr{\delta}_i$, with $i$ being the constant number, are listed in table~\ref{table:dragcoefficient-zero}.}\label{table:dragcoefficient-summary}
  \end{table}

\subsection{Lift coefficient\label{sec:liftcoeff}}

To ensure a comprehensive correlation to predict the lift coefficient of a particle at varying distances to the wall, particularly when the particle is situated outside the boundary layer, PR-DNS results from~\citet{Cheron2024} for the case of an isolated fixed rod-like particle subjected to a uniform fluid flow are used in the limit for large values of dimensionless distances $\tilde{L}$.
When the PR-DNS results are not available for the desired flow case, the correlation to predict the lift coefficient in case of uniform flow is used~\citep{Cheron2024}.
Although the accuracy of the correlation is not assessed for particles with aspect ratios lower than $\alpha < 2.5$, the resulots from the correlation to predict the lift coefficient of a sphere in case of a wall-bounded linear shear flow of~\citet{Zeng2009} are used as a limit.

\subsubsection{The effect of the distance to the wall on the lift coefficient of a particle with a fixed orientation angle} 
\begin{figure}[ht!]
    \centering
    \includegraphics[width=0.4\columnwidth, trim={0 0 0 0}, clip]{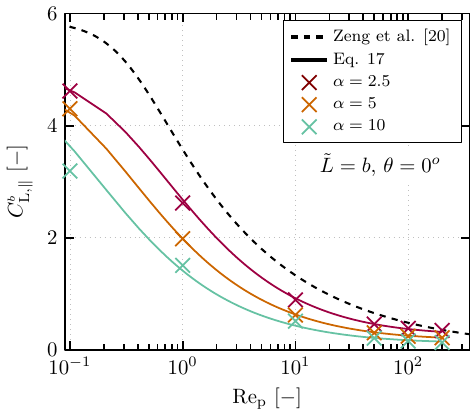}
    \caption{Evolution of the lift coefficient for a particle fixed to the wall at a fixed orientation angle of $\theta = 0^{o}$, \clpl, as a function of the particle Reynolds number, \Rep. Markers : PR-DNS results, Solid line: prediction given by Eq.~\ref{eq:cl-wall}. The colour indicates the aspect ratio of the particle, $\alpha = 10$ : \textcolor{blue!40!green!80}{green marker}, $\alpha = 5$ : \textcolor{black!20!orange}{orange marker}, $\alpha = 2.5$ : \textcolor{red!50!black}{red marker}. Results for a spherical particle are shown with dashed line: ~\citet{Zeng2009}.}
    \label{fig:liftcoefficient-evolution-Rep}
\end{figure}

To investigate the impact of wall distance on the lift coefficient of rod-like particles, insights from studies on spherical particles subject to a wall-bounded linear shear flow are used. Specifically, from the lubrication analysis of~\citet{Leighton1985} in the viscous regime, and the analysis by~\citet{Zeng2009} based on their PR-DNS results across varying particle Reynolds numbers.
In the former, the lubrication analysis of \citet{Leighton1985} provides a finite value for the lift coefficient of a spherical particle fixed to a wall of $5.87$.
From the PR-DNS simulations of a similar flow configuration,~\citet{Zeng2009} propose an expression to model the evolution of the lift coefficient of a spherical particle attached to the wall as a function of the particle Reynolds number, given by
\begin{equation}\label{eq:cl-zeng}
\text{\cll} = \frac{3.663}{\left(\text{\Rep}^2 + 0.1173\right)^{0.22}} .
\end{equation}
The prediction of the lift coefficient using this expression is shown in Figure \ref{fig:liftcoefficient-evolution-Rep}. 
The present PR-DNS results for a non-spherical rod-like particle with aspect ratios $\alpha = 2.5, 5$ and 10, at fixed orientation angle of $\theta = 0^{o}$, are also shown in this figure.\\

For all particle Reynolds numbers, the more elongated the particle, the lower is the value of the magnitude of the lift coefficient. Also, the evolution of the decrease in the magnitude of the lift coefficient as a function of the increase in the particle Reynolds number follows the trend of the correlation proposed by~\citet{Zeng2009} for spherical particles. This is also observed by~\citet{Fillingham2021} for prolate spheroids, although on a narrower range of particle Reynolds numbers, and for less elongated particles.
To include a larger range of particle Reynolds numbers, and include more elongated particles, we propose to model the lift coefficient for a particle in a shear flow fixed to the wall with an orientation angle of $\theta = 0^{o}$ as
\begin{equation}\label{eq:cl-wall}
\text{\clpb} = \left[\left(\frac{1}{1 + \text{\Rep}\alpha}\right)\left[\left(\left(\frac{1}{1 + \mathcal{\lambda}_1\text{\Rep}\alpha}\right) + \text{\Rep} \mathcal{\lambda}_2 + \mathcal{\lambda}_3 \right) \right]\right]^{\mathcal{\lambda}_4}\, ,\quad \text{valid for } \tilde{L} = b\, ,
\end{equation}
where the fit parameters $\lambda$ are listed in table~\ref{table:liftcoefficient-zero} under the appropriate equation number.

\begin{figure}[ht!]
    \centering
    \includegraphics[width=0.995\columnwidth, trim={0 0 0 0}, clip]{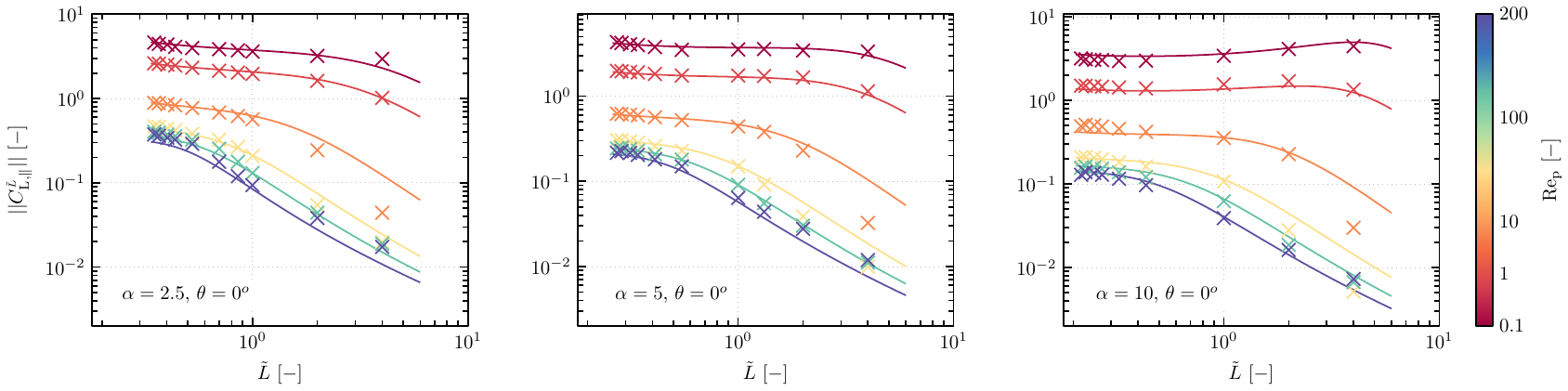}
    \caption{Evolution of the lift coefficient for a particle, \cl, at a fixed orientation angle of $\theta = 0^{o}$, as a function of the distance to the wall, $\tilde{L}$, for different aspect ratios, $\alpha$, and particle Reynolds numbers, \Rep. From left column to right column, $\alpha = 2.5$, $5$, and $10$. The colour map indicates the particle Reynolds number, \Rep. The solid line indicates the prediction given by Eq~\ref{eq:modellift-zerodegrees}.}
    \label{fig:liftcoefficient-evolution-Distance}
\end{figure}

\begin{figure}[ht!]
    \centering
    \includegraphics[width=0.64\columnwidth, trim={100 180 200 200}, clip]{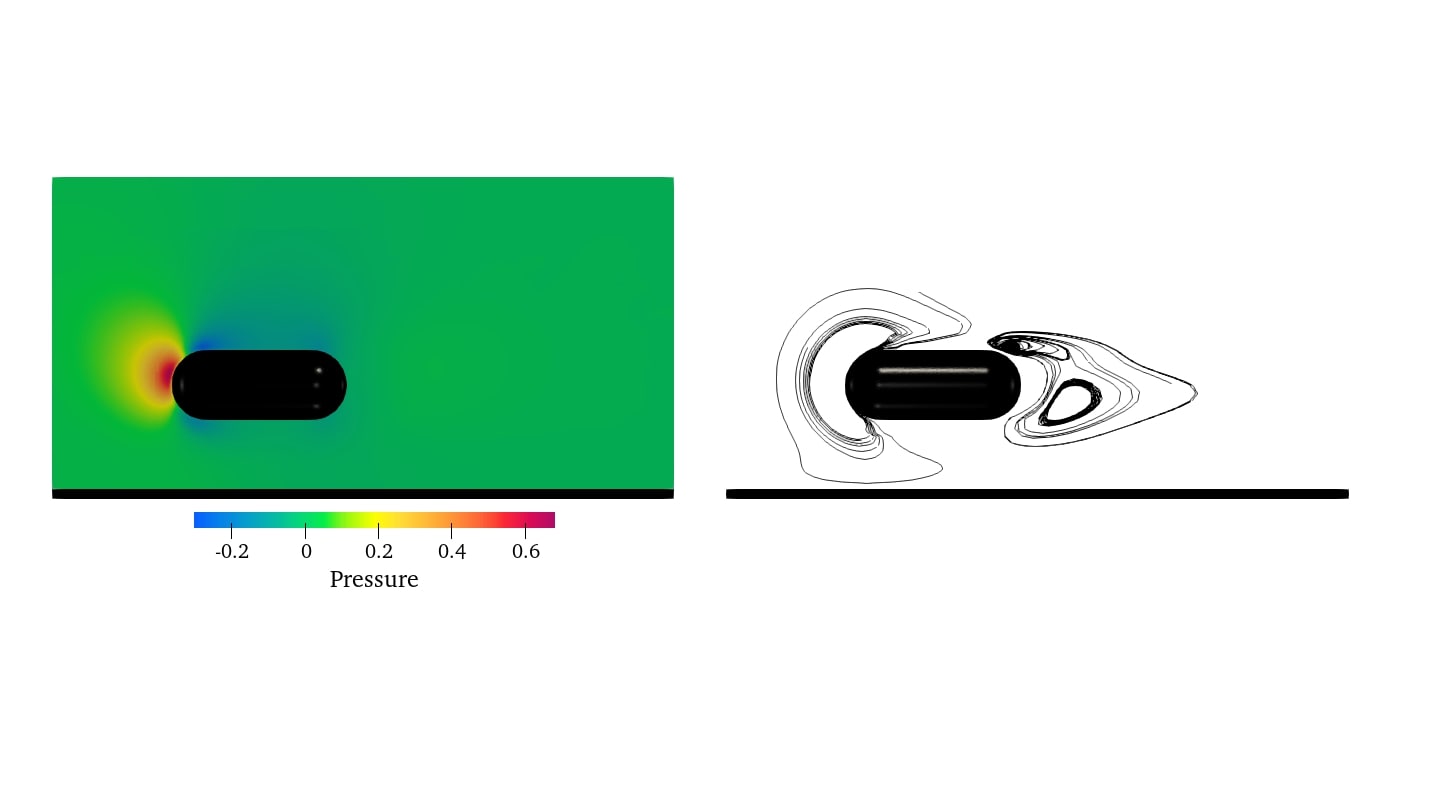}
    \caption{Snapshot of the pressure contours (left figure) and contours of fluid vorticity (right figure), resulting from the flow (left to right) past a fixed rod-like particle of aspect ratio $\alpha = 2.5$, at particle Reynolds number \Rep = 100, with orientation angle $\theta = 0^{o}$, for a dimensionless distance from the wall of $\tilde{L} = 1$.}
    \label{fig:flow-evolution-bis}
\end{figure}

The evolution of the magnitude of the lift coefficient as a function of the dimensionless distance to the wall is shown in figure~\ref{fig:liftcoefficient-evolution-Distance} for a particle with various aspect ratios and with a fixed orientation angle of $\theta = 0^{o}$, \clpl.
For this orientation angle, the presence of the wall-bounded linear shear flow profile yields to a fluid velocity recirculation between the wall and the particle, modifying the velocity profile in the wake of the particle.
This is illustrated in figure~\ref{fig:flow-evolution-bis} with a snapshot of the PR-DNS results for a particle with aspect ratio $\alpha = 2.5$, at particle Reynolds number \Rep = 100, dimensionless distance to the wall $\tilde{L} = 1$, and orientation angle set to $\theta = 0^{o}$.
From this figure, it is clear that the presence of the wall combined with the shear flow profile breaks the symmetry of the fluid flow profile past the particle, which induces a lift force.\\

The results shown in figure~\ref{fig:liftcoefficient-evolution-Distance} show a trend where an increase in particle Reynolds number corresponds
to a smaller absolute value of lift coefficient. 
This observation agrees with earlier findings, as larger particle Reynolds numbers generally coincide with a smaller magnitude of the lift coefficient across all orientation angles for non-spherical particles subjected to both uniform and non-uniform flows \citep{Zastawny2012c,Ouchene2016,Sanjeevi2018,Frohlich2020,Cheron2024}. Furthermore, the results shown in Figure \ref{fig:liftcoefficient-evolution-Distance} demonstrate the influence of particle Reynolds number on the decrease of the magnitude of the lift coefficient with respect to the dimensionless distance from the wall. For particle Reynolds numbers greater than or equal to 10, and all aspect ratios, the magnitude of the lift coefficient exhibits a one-order-of-magnitude decrease from $\tilde{L} = b$ to $\tilde{L} = 4$. Similar observations are reported for spherical particles in the study by~\citet{Zeng2009}.\\

For small particle Reynolds numbers, the evolution of the lift coefficient as a function of the distance to the wall does not follow the same trend.
For example, at a particle Reynolds number of \Rep = 0.1, the  lift coefficient as a function of the dimensionless distance from the wall remains nearly constant. Furthermore, for a particle with an aspect ratio of $\alpha = 10$, the magnitude of the lift coefficient shows a slightly larger value at a dimensionless distance from the wall of $\tilde{L} = 4$ compared to $\tilde{L} = b$. This suggests that in the viscous regime, the presence of the wall has a much smaller effect compared to the shear-induced lift force for elongated particles with orientation angle $\theta = 0^{o}$.\\

The proposed correlation to predict the lift coefficient of a particle with a fixed orientation angle, set to $\theta = 0^{o}$, varying as a function of the aspect ratio, the distance to the wall, and the particle Reynolds number, is given by
\begin{equation}\label{eq:modellift-zerodegrees}
    \text{\clpl} = \text{\clpb}\left[\frac{1}{1+X}\right]\, ,  \text{with } X = \frac{\left(1.025\tilde{L} - b\right)}{b^{-1}\left[\left(\left(\left[(\text{\Rep}(1.025\tilde{L} - b)) - (\alpha - \tilde{L})\right]\mathcal{\lambda}_1\right) + b\right) + \tilde{L}^{-1}\right]^{-1} + \mathcal{\lambda}_2}\, ,
\end{equation}
where the fit parameters $\lambda$ are listed in table~\ref{table:liftcoefficient-zero} under the appropriate equation number.
The prediction of the lift coefficients given by Eq.~\eqref{eq:modellift-zerodegrees}, for all particles, are shown in figure~\ref{fig:liftcoefficient-evolution-Distance}.
With this correlation, the effects of wall-bounded linear shear profile are well captured in the viscous regime for particles at varying distances to the wall, as well as the strong decay of the lift coefficient as a function of the distance to the wall for high particle Reynolds number cases.
For particle Reynolds number \Rep = 0.1, and a particle with aspect ratio $\alpha = 10$, there is still a limitation in the prediction of the lift coefficient for distances larger than $\tilde{L} \geq 10$, a correlation specifically developed for unbounded linear shear flow for this specific orientation angle~\citep{Cheron2024} will likely be more accurate.

\subsubsection{Analysis of the effect of the orientation of the near-wall particle on the lift coefficient}

\citet{Brenner1963} demonstrates that for an axi-symmetric non-spherical particle in a locally non-inertial uniform flow, the change of the lift coefficient with respect to the orientation angle can be expressed in terms of the drag coefficient only. 
This relationship is given by
 \begin{equation}
    \text{\cl} = \left[\text{\cdpp} - \text{\cdp}\right]\cos(\theta)\sin(\theta)\, .
 \end{equation} 
 The form of this expression is used by to build several correlations to predict the lift coefficient in case of uniform flow at finite particle Reynolds number~\citep{Frohlich2020,Sanjeevi2022,Cheron2024}.
 In case of unbounded linear shear flow,~\citet{Harper1968} propose to model the shear-induced lift force by including an additional force to the lift force in case of uniform flow.
 This force acts from the lower to the higher fluid velocity regardless of the orientation of the particle.
 Thus, it modifies the profile of the evolution of the lift coefficient as a function of the orientation angle. 
 From the results of~\citet{Cheron2024}, it is shown that as the particle Reynolds number increases, the change in the value of the lift coefficient in case of unbounded linear shear flow, compared to uniform flow, decreases until becoming very small.\\
 
\begin{figure}[ht!]
    \centering
    \begin{tabular}{m{0.03\linewidth}m{0.95\linewidth}}
    \rotatebox{90}{$\quad$ \Rep = 0.1} &
    \includegraphics[width=0.95\columnwidth, trim={0 0 0 0}, clip]{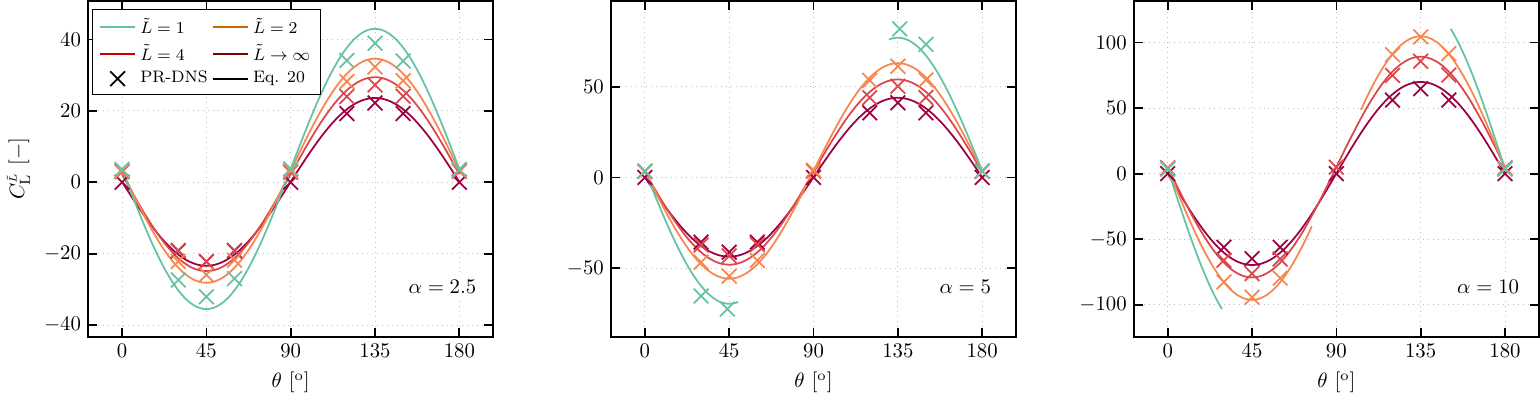}\\
    \rotatebox{90}{$\quad$ \Rep = 10} &
    \includegraphics[width=0.95\columnwidth, trim={0 0 0 0}, clip]{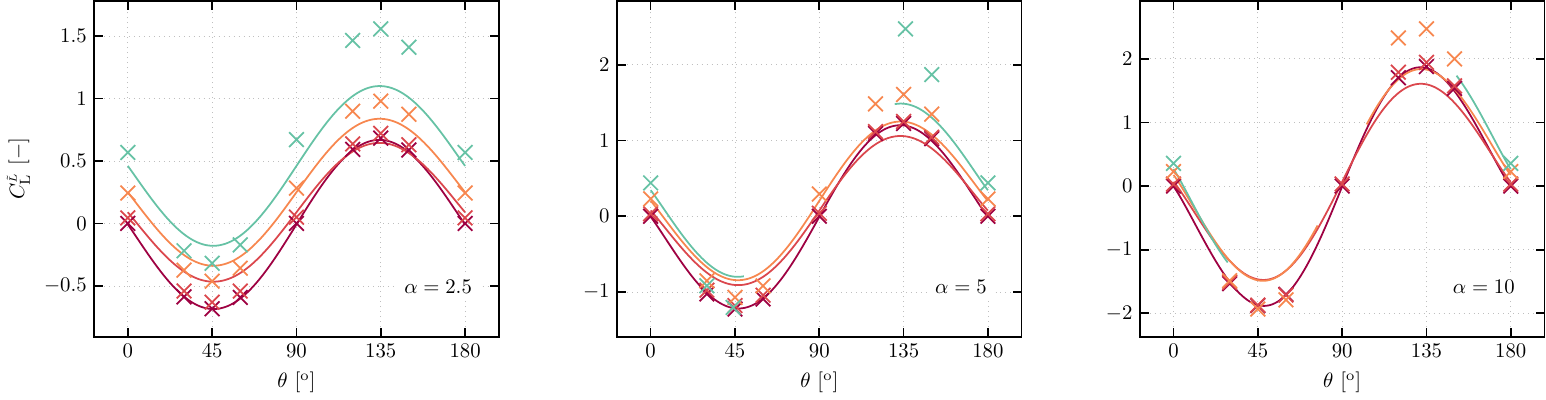}\\
    \rotatebox{90}{$\quad$ \Rep = 200} &
    \includegraphics[width=0.95\columnwidth, trim={0 0 0 0}, clip]{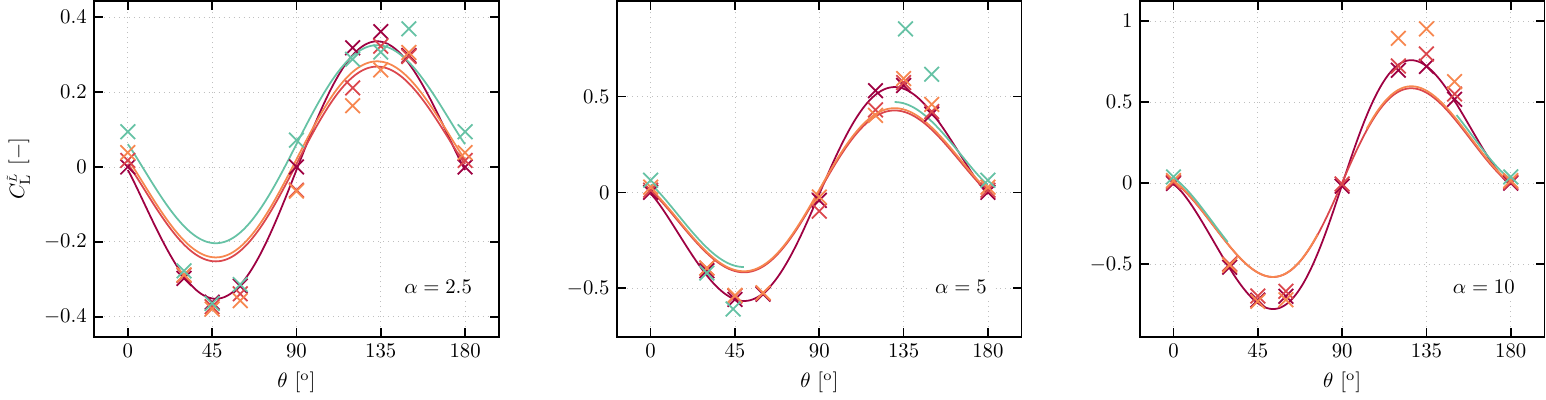}\\
    \end{tabular}
     \caption{Evolution of lift coefficient, \cl, as a function of the orientation angle, $\theta$. From left column to right column, $\alpha = 2.5$, $5$, and $10$.  From top row to bottom row, \Rep = 0.1, 10 and 200. Marker indicates the PR-DNS results, solid line indicates the novel correlation. Colour indicates distance to the wall, $\tilde{L} = \infty$ : \textcolor{black!50!red}{dark red}, $\tilde{L} = 4$ : \textcolor{black!20!red}{red}, $\tilde{L} = 2$ : \textcolor{black!20!orange}{orange}, $\tilde{L} = 1$ : \textcolor{blue!40!green}{green}.}
    \label{fig:liftcoefficient-fit}
\end{figure}

To study the influence of the orientation angle on the lift coefficient in case of wall-bounded linear shear flow, the evolution of the lift coefficient relative to the lift coefficient in the case of uniform flow, as a function of the orientation angle, is shown in figure \ref{fig:liftcoefficient-fit}.
The results are shown as a function of the orientation angle for the particle Reynolds numbers of \Rep = 0.1, 1, and 100, and the dimensionless distances from the wall of $\tilde{L} = 1$, 2 and 4, and the particles with aspect ratios of $\alpha$ = 2.5, 5 and 10.
At a particle Reynolds number of \Rep = 0.1, a flow configuration with an unbounded-linear shear flow increases the magnitude of the lift coefficient. This additional force acts from the lower to the higher fluid velocity regardless of the orientation of the particle~\citep{Harper1968}.
However, in this study we observe that the
orientation of the particle relative to the flow, combined with the wall distance, affects the increase in the magnitude of the lift force coiefficient.
For instance, for all particles considered in the range of orientation angles from $\theta = 90^o$ to $180^o$, the additional lift force contribution caused by the wall-bounded linear shear flow acts from the lower fluid velocity to the higher fluid velocity,  whereas in the range of orientation angles from $\theta = 0^o$ to $90^o$, the additional lift force contribution caused by the wall-bounded linear shear flow acts from the higher fluid velocity towards the lower fluid velocity.
In addition, the closer the particle is to the wall, the higher is the influence of the orientation angle on the change of the lift coefficient.\\

As the particle Reynolds number increases, the additional contribution to the lift force caused by the wall-bounded linear shear flow profile always increases for all particle aspect ratios and their distances to the wall considered in this work. However, the influence of the orientation angle on the evolution of the lift force is more challenging to capture.
For a particle with an aspect ratio of $\alpha =2.5$, the closer the particle is to the wall, the larger is the increase in the magnitude of the lift force coefficient.
However, for particles with a larger aspect ratio, we observe that the change in the magnitude of the lift force coefficient in the range of orientation angles $\theta = 0^{o}$ to $90^{o}$ is less significant.
The PR-DNS results almost overlap with the results in case of uniform flow, for both closest and furthest positions. This is observed for the PR-DNS results for the particle with aspect ratio $\alpha =5$, with an orientation angle $\theta = 45^{o}$.\\

At particle Reynolds number \Rep = 200, the change in the magnitude of the lift coefficient compared to the lift coefficient in case of uniform flow drastically reduces for all aspect ratios, and is rather complex to describe, as it is drastically affected by the flow recirculation between the wall and the particle, see figure~\ref{fig:flow-evolution}. For instance, we observe for particles with aspect ratios $\alpha = 2.5$ and 5, and with an orientation angle of $\theta = 90^{o}$, a negative lift coefficient for intermediate distance to the wall, $\tilde{L} \geq 2$, as in the case of unbounded linear shear flow, meanwhile this value remains positive for the closest distance to the wall $\tilde{L} \leq 1$.
The interaction between the shear profile and wall effects is challenging to accurately describe. As a solution, we propose a correlation designed to best represent these dynamics, albeit with a trade-off in absolute precision that could, perhaps, be achieved through a piecewise approach.\\

The correlation to predict the evolution of the lift coefficient of a particle with a fixed orientation angle of $\theta = 0^{o}$, Eq.~\eqref{eq:modellift-zerodegrees}, is used as a based function to provide our general model that varies as a function of the orientation angle, distance to the wall, particle Reynolds number, and aspect ratio. This expression is given by 
\begin{equation}\label{eq:fullcl}
\text{\cll} = \text{\clpl} + \left[\left(\frac{\mathscr{\lambda}_1}{(\text{\Rep} +  \mathscr{\lambda}_2)^{\left(1/b\right)}} - \frac{\alpha/\text{\Rep}}{1+((1.025\tilde{L} - b)^{ \left(\mathscr{\lambda}_3\right)})}\right) - \frac{\mathscr{\lambda}_4}{\text{ln}(1.025\tilde{L} - b) - \text{\Rep}}\right] 2\cos(\theta)\sin(\theta)\, ,
\end{equation}
where the fit parameters $\lambda$ are listed in table~\ref{table:liftcoefficient-zero} under the appropriate equation number.\\

\begin{table}
    \centering
    \begin{tabular}{l | c c c c c c c c c c c}
     & $\mathscr{\lambda}_1$ & $\mathscr{\lambda}_2$ & $\mathscr{\lambda}_3$ & $\mathscr{\lambda}_4$ \\
    \hline
    \hline
    Eq.~\ref{eq:cl-wall} & -1.244 & 0.131 & 22.567 & 0.518 \\
    Eq.~\ref{eq:modellift-zerodegrees} & 0.146 & 0.106 & - & - \\
    Eq.~\ref{eq:fullcl} & $-4.084 \times 10^{-8}$ & $7.990 \times 10^{-2}$ & 1.252 & -0.196 \\

\end{tabular}
\caption{List of the fit parameters in Eqs.\ref{eq:cl-wall},~\ref{eq:modellift-zerodegrees} and~\ref{eq:fullcl} used in the correlation to predict the change in the lift coefficient in case of wall-bounded linear shear flow, with respect to the uniform flow lift coefficient.}\label{table:liftcoefficient-zero}
\end{table}

The correlation derived to predict the change in the lift coefficient in case of wall-bounded linear shear flow compared to the uniform flow coefficient, is shown in figure~\ref{fig:torquecoefficient-fit}.
Although the change in the lift coefficient for a particle with an orientation anle of $\theta = 90^{o}$ is not captured by the prediction, in general, a good agreement between the PR-DNS results and the model fit is observed, and more especially for low particle Reynolds number.
For instance, a median relative error between the model prediction and the PR-DNS results of ${\mathcal{E}}$ = 5.37\% is obtained.
The correlation coefficient between the model prediction and the results is equal to \color{black}{$\mathcal{R}^{2}$ = 0.99}.
This correlation can predict the lift coefficient for a particle with an aspect ratio ranging from $\alpha = 2.5$ to 10, particle Reynolds numbers ranging from \Rep = 0.1 to 200, orientation angles from $\theta = 0^{o}$ to $180^{o}$ (where the angle varies in the fluid flow direction and wall-normal direction plane), and dimensionless distances from the wall in the range $\tilde{L} = b$ to $\tilde{L} =\infty$, where $\tilde{L}\rightarrow\infty$ implies locally uniform flow conditions.

\subsubsection{Summary of lift coefficient correlation and range of validity\label{sec:lift-correl}}

The correlation derived to predict the change in the lift coefficient of a particle in case of wall-bounded linear shear flow compared to the lift coefficient in case of a uniform flow, is summarized in table~\ref{table:liftcoefficient-summary}.
This correlation depends on the particle Reynolds number, \Rep, the orientation angle between the particle major axis and the local direction of the fluid flow, $\theta$, the aspect ratio of the particle, $\alpha$, and the dimensionless distance between the centre of the particle and the wall $\tilde{L}$.
The correlation to predict the change in the lift coefficient in case of wall-bounded linear shear flow is valid for particle Reynolds numbers in the range of 0.1 $\leq$ \Rep  $\leq$ 200, orientation angles in the range  of $0^{o}\leq \theta \leq 180^{o}$, where the orientation angle varies in the fluid flow direction and wall-normal direction plane, particles with an aspect ratio in the range of $2.5 \leq \alpha \leq 10$, and dimensionless distance from touching the wall, $\tilde{L} = b$, to outside the boundary layer $\tilde{L} \rightarrow \infty$. The latter condition ensures that as $\tilde{L}$ increases, the present correlation asymptotically approaches zero, so that the lift force for a uniform flow coefficient is achieved outside the wall boundary layer. 

\begin{table}
  \centering
  \resizebox{\columnwidth}{!}{%
  \begin{tabular}{l | r c l l}
     & Coefficient &  & Formula & \\
    \hline
    \hline
    {Eq.~\eqref{eq:fullcl}} & {\cll - \cl} & {=} & {$ \text{\clpl} + \left[\left(\frac{\mathscr{\lambda}_1}{(\text{\Rep} +  \mathscr{\lambda}_2)^{\left(1/b\right)}} - \frac{\alpha/\text{\Rep}}{1+((1.025\tilde{L} - b)^{ \left(\mathscr{\lambda}_3\right)})}\right) - \frac{\mathscr{\lambda}_4}{\text{ln}(1.025\tilde{L} - b) - \text{\Rep}}\right] 2\cos(\theta)\sin(\theta) $ }& {}\\[4mm]
    {Eq.~\eqref{eq:modellift-zerodegrees} } & {\clpl} & {=} & {$ \text{\clpb}\left[\frac{1}{1+X\left(\text{\Rep},\alpha,\tilde{L},b\right)}\right]$}&\\
    - &  $X\left(\text{\Rep},\alpha,\tilde{L},b\right)$ & = & ${\left(1.025\tilde{L} - b\right)}\left[b^{-1}\left[\left(\left(\left[(\text{\Rep}(1.025\tilde{L} - b)) - (\alpha - \tilde{L})\right]\mathcal{\lambda}_1\right) + b\right) + \tilde{L}^{-1}\right]^{-1} + \mathcal{\lambda}_2\right]^{-1}$& {for $\theta = 0^{o}$}\\[4mm]
    {Eq.~\eqref{eq:cl-wall} } & {\clpb}  & {=} & {$ \left[\left(\frac{1}{1 + \text{\Rep}\alpha}\right)\left[\left(\left(\frac{1}{1 + \mathcal{\lambda}_1\text{\Rep}\alpha}\right) + \text{\Rep} \mathcal{\lambda}_2 + \mathcal{\lambda}_3 \right) \right]\right]^{\mathcal{\lambda}_4}$ }& {for $\tilde{L} = b$ and $\theta = 0^{o}$}\\[4mm]
    \hline\\
    {~\citep[Eq. 40]{Cheron2024}} & {\cl}  & {}& {\citet{Cheron2024}} & {}\\[1mm]
  \end{tabular}
  }
       \caption{Correlation to predict the lift coefficient of a particle subjected to a wall-bounded linear shear flow as a function of the particle Reynolds number, \Rep, the orientation angle between the major axis of the particle and the local fluid flow direction, $\theta$, the aspect ratio of the particle, $\alpha$, and the dimensionless distance between the centre of the particle and the wall, $\tilde{L}$. The lift coefficient, {\cll}, is independent of the lift coefficient of a particle in case of uniform flow, {\cl}, and is given as a function of the coefficients of the same particle with an orientation angle fixed to $\theta = 0^{o}$, {\clpl}, for the same flow conditions. The fit parameters, $\mathcal{\lambda_i}$, with $i$ being the constant number, are listed in table~\ref{table:liftcoefficient-zero}.}\label{table:liftcoefficient-summary}
  \end{table}

\subsection{Torque coefficient\label{sec:torquecoeff}}

To ensure that the correlations derived in this section to predict the torque coefficient in case of wall-bounded linear shear flow are continuous from inside the wall-boundary layer to outside the wall boundary layer, the PR-DNS results from~\citet{Cheron2024} for the case of an isolated fixed rod-like particle subject to a uniform fluid flow are incorporated to the set of data, when available, for the model, setting the limits of the model as $\tilde{L} \rightarrow \infty$.  Otherwise, the values are obtained from the torque correlation for the case of uniform flow~\citep{Cheron2024}.

\subsubsection{Analysis of the effect of the distance to the wall on the torque coefficient of a particle with fixed orientation angle} 

\begin{figure}[ht!]
    \centering
    \includegraphics[width=0.4\columnwidth, trim={0 0 0 0}, clip]{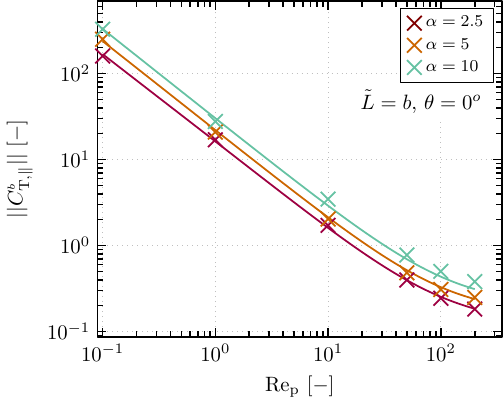}
    \caption{Evolution of the torque coefficient for a particle fixed to the wall at a fixed orientation angle of $\theta = 0^{o}$, \ctpb, as a function of the particle Reynolds number, \Rep. The colour indicates the aspect ratio of the particle, $\alpha = 10$ : \textcolor{blue!40!green!80}{green}, $\alpha = 5$ : \textcolor{black!20!orange}{orange}, $\alpha = 2.5$ : \textcolor{red!50!black}{red}. Markers are PR-DNS result, and solid lines the prediction given by Eq.~\eqref{eq:torque-fixed-wall-anglezero}.}
    \label{fig:torquecoefficient-evolution-Rep}
\end{figure}

The evolution of the absolute value of the torque coefficient for a particle fixed to the wall is shown in figure~\ref{fig:torquecoefficient-evolution-Rep} as a function of the particle Reynolds number.
The PR-DNS results for all the particles are shown for an orientation angle of the particle fixed to $\theta = 0^{o}$.
For this specific orientation angle, the value of the torque coefficient in case of uniform flow is always equal to zero, \ctp = 0, \citep{Zastawny2012c,Sanjeevi2018,Cheron2024}.
For a particle fixed to the wall with an orientation angle of $\theta = 0^{o}$, the magnitude of the torque coefficient of all particles for all aspect ratios decreases logarithmically as a function of the particle Reynolds.
For large particle Reynolds numbers, however, the slope of the decrease is less pronounced, which is observed for all particles.
The evolution of the torque coefficient for a particle fixed to the wall with an orientation angle of $\theta = 0^{o}$ is predicted with
\begin{equation}\label{eq:torque-fixed-wall-anglezero}
    \text{\ctpb} = \frac{\zeta_1\text{\Rep}^{\zeta_2}}{\alpha^{\zeta_3}} + \frac{\zeta_4 \text{\Rep}^{\zeta_5}}{\alpha^{\zeta_6}} \, ,\quad \text{valid for } \tilde{L} = b\, ,
    \end{equation}
where the fit parameters $\zeta$ are listed in table~\ref{table:torquecoefficient-zero} under the appropriate equation number.

The prediction of the torque coefficient given by Eq.~\eqref{eq:torque-fixed-wall-anglezero} is also shown in figure~\ref{fig:torquecoefficient-evolution-Rep}, and a very good agreement is observed between the PR-DNS results and the model fit.

\begin{figure}[ht!]
    \centering
    \includegraphics[width=0.995\columnwidth, trim={0 0 0 0}, clip]{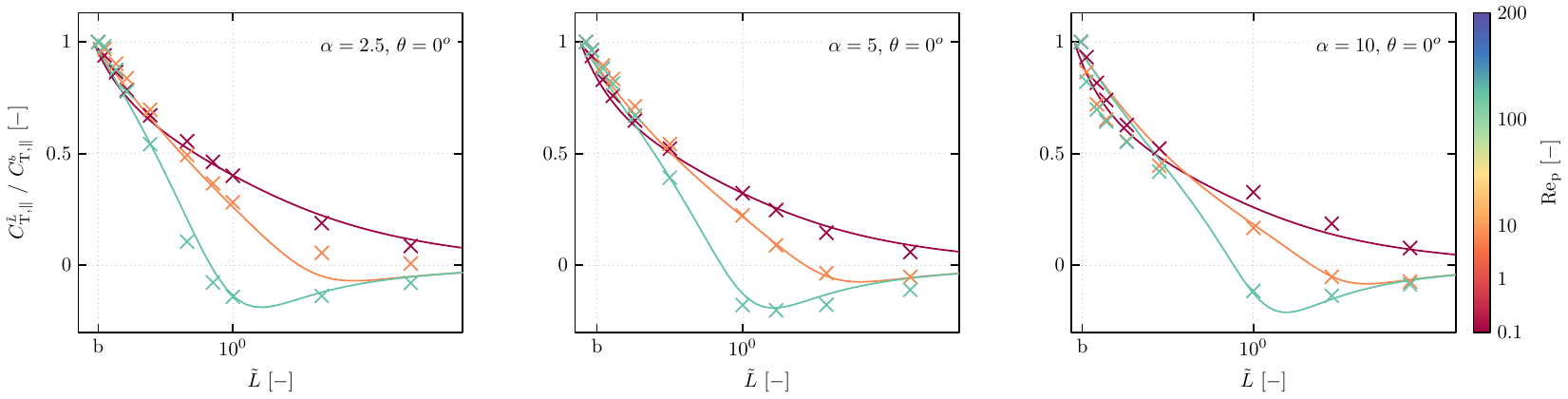}
    \caption{Evolution of the torque coefficient at a fixed orientation angle of $\theta = 0^{o}$, scaled by the torque coefficient of a particle fixed to the wall, as a function of the distance to the wall. From left figure to right figure, $\alpha = 2.5$, $5$, and $10$. The colour map indicates the particle Reynolds number, \Rep = 0.1, 10 and 100. The solid lines indicate the prediction given by Eq.~\ref{eq:torque-distance-anglezero}.}
    \label{fig:torquecoefficient-evolution-Distance}
\end{figure}

To assess the change in the torque coefficient for a particle subject to a wall-bounded linear shear flow at varying distances from the wall, the torque coefficient evolution is shown in figure~\ref{fig:torquecoefficient-evolution-Distance}, as a function of the distance to the wall.
The torque coefficient is scaled by the torque coefficient of a particle fixed to the wall, given by Eq.~\ref{eq:torque-fixed-wall-anglezero}.
The results are shown for the particle Reynolds numbers \Rep = 0.1, 10 and 100, and the orientation of the particle remains equal to~$\theta = 0^{o}$.

The further the particle is from the wall, the closer the torque coefficient is to the torque coefficient of a particle subject to a locally uniform flow, \textit{i.e.,} \ctp = 0.
The particle Reynolds number influences the rate of decay of the evolution of the torque coefficient as a function of the distance to the wall.
The larger the value of the particle Reynolds number, the faster the torque coefficient of the particle decays. 
Also, at large particle Reynolds number, \Rep = 100, the torque coefficient of the particle changes of sign for dimensionless distances larger than $\tilde{L} = 0.8$.
This is observed for all the particles studied in this work, and is caused by the specific fluid flow recirculation in the wake of the particle, as illustrated in figure~\ref{fig:flow-evolution-bis}.
This effect vanish for the largest dimensionless distances to the wall studied in this work, and the torque coefficient reduces to the values in case of a locally uniform flow.
The influence of the distance to the wall on the prediction of the torque coefficient of a particle parallel to the wall can be modelled based on the correlation for a particle fixed to the wall, Eq.~\ref{eq:torque-fixed-wall-anglezero}, given by
\begin{equation}\label{eq:torque-distance-anglezero}
    \frac{\text{\ctpl}}{\text{\ctpb}} = \frac{1}{1+\left(\tilde{L} - b\right) \left[\left(\frac{\frac{\tilde{L}^{-1}}{\text{\Rep}} + \zeta_{1} }{\frac{\zeta_{2}}{\text{\Rep}^{\tilde{L}}} - \zeta_{3}}\right) - \frac{\zeta_{4}}{b}\right]}\frac{1}{\zeta_5}\ ,
\end{equation}
where the fit parameters $\zeta$ are listed in table~\ref{table:torquecoefficient-zero} under the appropriate equation number.

The prediction of the torque coefficient given by Eq.~\ref{eq:torque-distance-anglezero} is also shown in figure~\ref{fig:torquecoefficient-evolution-Distance}, a very good agreement is observed between the PR-DNS results and the model fit.

\subsubsection{Analysis of the effect of the orientation of the near-wall particle on the torque coefficient}

\begin{figure}[ht!]
    
    \centering
    \begin{tabular}{m{0.03\linewidth}m{0.95\linewidth}}
        \rotatebox{90}{$\quad$ \Rep = 0.1} &
         \includegraphics[width=0.95\columnwidth, trim={0 0 0 0}, clip]{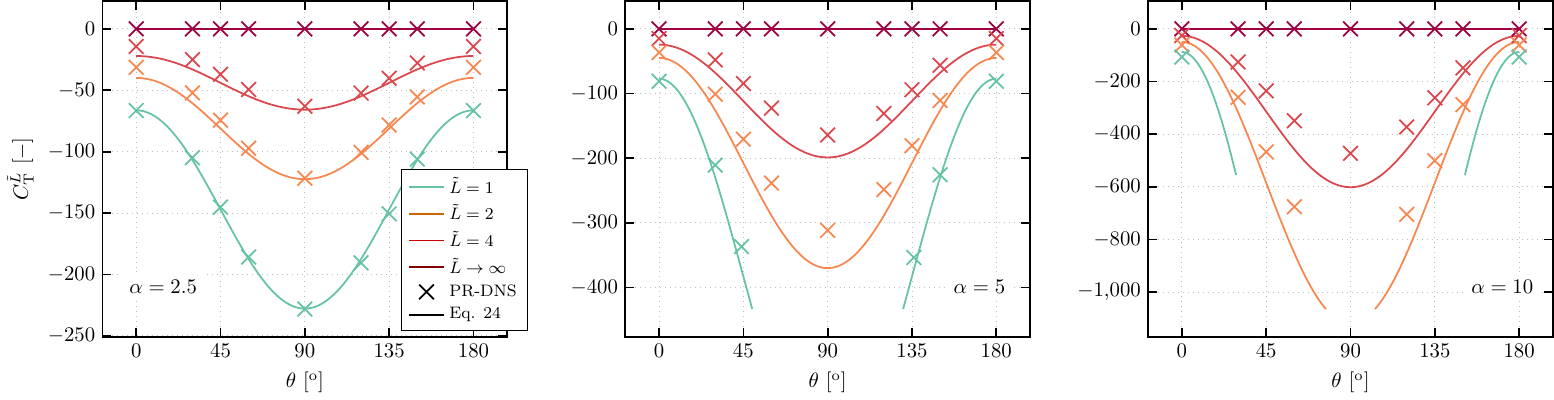}\\
        \rotatebox{90}{$\quad$ \Rep = 10} &
         \includegraphics[width=0.95\columnwidth, trim={0 0 0 0}, clip]{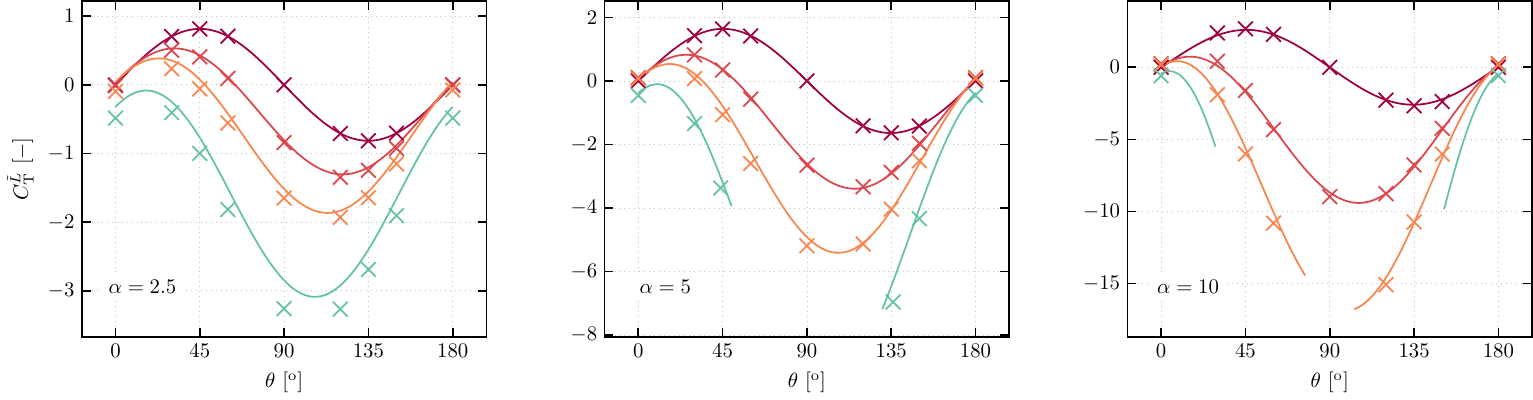}\\
        \rotatebox{90}{$\quad$ \Rep = 200} &
         \includegraphics[width=0.95\columnwidth, trim={0 0 0 0}, clip]{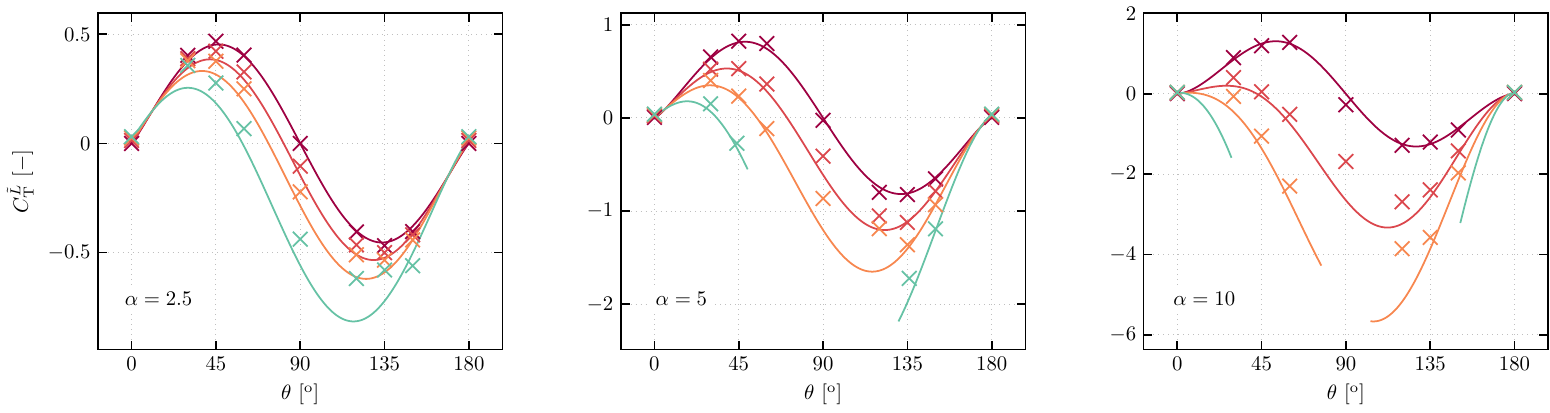}\\
    \end{tabular}
     \caption{Evolution of the torque coefficient, \ct, as a function of the orientation angle, $\theta$. From left column to right column, $\alpha = 2.5$, $5$, and $10$.  From top row to bottom row, \Rep = 0.1, 10 and 200. Colour indicates distance to the wall, $\tilde{L} = \infty$ : \textcolor{black!50!red}{dark red}, $\tilde{L} = 4$ : \textcolor{black!20!red}{red}, $\tilde{L} = 2$ : \textcolor{black!20!orange}{orange}, $\tilde{L} = 1$ : \textcolor{blue!40!green}{green}. PR-DNS results: markers, solid lines: prediction given by Eq.~\ref{eq:CT}.}
    \label{fig:torquecoefficient-fit}
\end{figure}

The analysis of the interactions forces on a rod-like particle in an unbounded linear shear flow of~\citet{Cheron2024} suggests that the change in the torque coefficient in case of a locally unbounded linear shear flow, in comparison to the case with a locally uniform flow, can be characterized by a ``sinesquare'' profile. Consequently, their analysis reveals that the minimum and maximum changes in the torque coefficient occur at $\theta = 0^{o}$ and $\theta = 90^{o}$, respectively.
The study of~\citet{Sanjeevi2018} shows that the torque coefficent of a non-spherical particle in a uniform flow in the viscous regime is equal to  \ct = 0. In this work, this is observed for a particle Reynolds number of \Rep = 0.1, and a dimensionless distance of $\tilde{L}\rightarrow\infty$. Thus, for a particle Reynolds number of \Rep = 0.1, the evolution of the torque coefficient as a function of the orientation angle, shown in figure 15, is equivalent to the evolution of the torque coefficient in case of wall-bounded linear shear flow velocity profile.
This is shown in the first row of figure~\ref{fig:torquecoefficient-fit} for all particles considered in this work and several dimensionless distances to the wall. 
The PR-DNS results for the flow regime of \Rep = 0.1 show that the largest change compared to the torque coefficient in case of uniform flow is always obtained at an orientation angle of $\theta = 90^{o}$. 
For all orientation angles, the more the particle is elongated, and the closer the particle is to the wall, the larger the change in the torque coefficient, compared to the case with uniform flow.
Also, the symmetry between the PR-DNS values at the orientation angle of $\theta = 90^{o}$ is not perfect, \textit{i.e.,} the torque coefficient of a particle with orientation angle in the range $\theta>90^{o}$ is slightly larger than for a particle with orientation angle in the range $\theta<90^{o}$, the assumption that the change in the torque coefficient as a function of the orientation angle evolves with a ``sinesquare'' profile is a good assumption for a model fit for low particle Reynolds number.

For larger particle Reynolds numbers, i.e., \Rep = 10, and 200, shown in the second and last rows of figure~\ref{fig:torquecoefficient-fit}, the change in the torque coefficient in case of wall-bounded linear shear flow compared to the case with uniform flow remains maximum for the particles with orientation angle of $\theta = 90^{o}$.
However, the symmetric behaviour of the evolution of the torque coefficient of the particle as a function of the orientation angle around $\theta = 90^{o}$, is lost for high particle Reynolds number and particles located close to the wall.
In addition, for all flow cases at \Rep = 200, the change in the torque coefficient is always greater for the cases with orientation angles ranging from $\theta = 0^{o}$ to $90^{o}$, compared to the orientation angles in the range $\theta = 90^{o}$ to $180^{o}$.
Thus, the evolution of the particle Reynolds number changes the range of orientation angle where the change is the largest.
Nevertheless, this deviation from the ideal ``sinesquare'' profile is not significant enough to deviate from the ``sinesquare'' model fit.
Thus, the correlation to predict the torque coefficient from a particle touching the wall to a particle outside the wall-boundary layer is modelled with the expression
\begin{equation}\label{eq:CT}
\text{\ctl} = \text{\ct} + \text{\ctpl} + \left[\text{\ctplp} - \text{\ctpl}\right]\sin^{2}\left(\theta\right)\, ,
\end{equation} 
where only the term \ctplp remains to be determined, and is given by
\begin{equation}\label{eq:torque-perp}
\text{\ctplp} = \zeta_{1} \alpha^{\zeta_{2}} \frac{\zeta_{3}}{\tilde{L}^{\zeta_{4}}} \zeta_{5} \frac{\tilde{L}}{\zeta_{6} + \zeta_{7}\text{\Rep}^{\zeta_{8}}} + \frac{\zeta_{9}\alpha^{\zeta_{10}}}{\tilde{L}}\, ,
\end{equation} 
where the fit parameters $\zeta$ are listed in table~\ref{table:torquecoefficient-zero} under the appropriate equation number.\\

The correlation derived to predict the change in the torque coefficient in case of wall-bounded linear shear flow compared to the torque coefficient in a uniform flow, is shown in figure~\ref{fig:torquecoefficient-fit}.
A generally good agreement between the PR-DNS results and the model fit is observed, and a median relative error between the model prediction and the PR-DNS results of ${\mathcal{E}}$ = 11.0 \% is obtained.
The correlation coefficient between the model prediction and the results is equal to $\mathcal{R}^{2}$ = 0.955.
This correlation can predict the torque coefficient for a particle with an aspect ratio ranging from $\alpha = 2.5$ to 10, particle Reynolds numbers ranging from \Rep = 0.1 to 200, orientation angles from $\theta = 0^{o}$ to $180^{o}$ (where the angle varies in the fluid flow direction and wall-normal direction plane), and dimensionless distances from the wall in the range $\tilde{L} = b$ to $\tilde{L} =\infty$, where $\tilde{L}\rightarrow\infty$ implies locally uniform flow conditions.

\begin{table}
    \centering
    \resizebox{\columnwidth}{!}{%
    \begin{tabular}{l || c c c c c c c c c c}
     & $\mathscr{\zeta}_1$ & $\mathscr{\zeta}_2$ & $\mathscr{\zeta}_3$ & $\mathscr{\zeta}_4$ & $\mathscr{\zeta}_5$ & $\mathscr{\zeta}_6$ & $\mathscr{\zeta}_7$ & $\mathscr{\zeta}_8$ & $\mathscr{\zeta}_9$ & $\mathscr{\zeta}_{10}$ \\
    \hline
    \hline
    Eq.~\eqref{eq:torque-fixed-wall-anglezero} & 2.141  & -1.024 & 1.292 & 7.4 $\times$ $10^{-3}$ & 0.253 & 0.878 & - & - & - & -  \\
    Eq.~\eqref{eq:torque-distance-anglezero} & 16.919 & 81.870 & 1.675 & -0.417 & 0.754 & - & - & - & - & -\\
    Eq.~\eqref{eq:torque-perp} & 1.807 & 1.595 & -417.657 & 1.896 & 0.0182 & -0.191 & 4.396 & 0.894 & -0.0498 & 2.249 \\
    \end{tabular}
    }
         \caption{List of the fit parameters in Eqs.~\eqref{eq:torque-fixed-wall-anglezero}, ~\eqref{eq:torque-distance-anglezero}, and~\eqref{eq:torque-perp}, used in the correlation to predict the change in the torque coefficient in case of wall-bounded linear shear flow, with respect to the torque coefficient in case of uniform flow, along with their accuracy expressed as a correlation coefficient and root-mean-square error.}\label{table:torquecoefficient-zero}
    \end{table}

\subsubsection{Summary of torque coefficient correlation and range of validity\label{sec:torque-correl}}

The correlation derived to predict the change in the torque coefficient in case of wall-bounded linear shear flow compared to the torque coefficient in case of uniform flow is summarized in the table~\ref{table:torquecoefficient-summary}.
This correlation varies as a function of the particle Reynolds number, \Rep, the orientation angle between the particle major axis and the local direction of the fluid flow, $\theta$, the aspect ratio of the particle, $\alpha$, and the dimensionless distance between the centre of the particle and the wall $\tilde{L}$.
The correlation to predict the change in the torque coefficient in case of wall-bounded linear shear flow is valid for particle Reynolds number in the range 0.1 $\leq$ \Rep  $\leq$ 200, orientation angle in the range  $0^{o}\leq \theta \leq 180^{o}$, where the orientation angle varies in the fluid flow direction and wall-normal direction plane, particle with an aspect ratio in the range $2.5 \leq \alpha \leq 10$, and dimensionless distance from touching the wall, $\tilde{L} = b$, to outside the boundary layer $\tilde{L} \rightarrow \infty$. The latter condition ensures that as $\tilde{L}$ increases, the present correlation asymptotically approaches zero.

\begin{table}
  \centering
  \begin{tabular}{l | r c l l}
     & Coefficient &  & Formula & \\
    \hline
    \hline
    {Eq.~\eqref{eq:fullcl}} & {\ctl - \ct} & {=} & {$  \text{\ctpl} + \left[\text{\ctplp} - \text{\ctpl}\right]\sin^{2}\left(\theta\right) $ }& {}\\[4mm]
    Eq.~\eqref{eq:torque-perp} & {\ctplp} & {=} & {$  \zeta_{1} \alpha^{\zeta_{2}} \frac{\zeta_{3}}{\tilde{L}^{\zeta_{4}}} \zeta_{5} \frac{\tilde{L}}{\zeta_{6} + \zeta_{7}\text{\Rep}^{\zeta_{8}}} + \frac{\zeta_{9}\alpha^{\zeta_{10}}}{\tilde{L}}$ }& {for $\theta = 90^{o}$}\\[4mm]
    Eq.~\eqref{eq:torque-distance-anglezero} & {\ctpl} & {=} & {$ \frac{1}{1+\left(\tilde{L} - b\right) \left[\left(\frac{\frac{\tilde{L}^{-1}}{\text{\Rep}} + \zeta_{1} }{\frac{\zeta_{2}}{\text{\Rep}^{\tilde{L}}} - \zeta_{3}}\right) - \frac{\zeta_{4}}{b}\right]}\frac{1}{\zeta_5}$ }& {for $\theta = 0^{o}$}\\[4mm]
    {Eq.~\eqref{eq:torque-fixed-wall-anglezero}} & \ctpb  & {=} & $\frac{\zeta_1\text{\Rep}^{\zeta_2}}{\alpha^{\zeta_3}} + \frac{\zeta_4 \text{\Rep}^{\zeta_5}}{\alpha^{\zeta_6}}$ & {for $\tilde{L} = b$ and $\theta = 0^{o}$}\\[4mm]
    \hline\\
    {~\citep[Eq. 51]{Cheron2024}} & {\ct}  & {}& {\citet{Cheron2024}} & {}\\[1mm]
  \end{tabular}
       \caption{Correlation to predict the torque coefficient of a particle subjected to a wall-bounded linear shear flow as a function of the particle Reynolds number, \Rep, the orientation angle between the major axis of the particle and the local fluid flow direction, $\theta$, the aspect ratio of the particle, $\alpha$, and the dimensionless distance between the centre of the particle and the wall, $\tilde{L}$. The torque coefficient, {\ctl}, is independent of the torque coefficient of a particle in case of uniform flow, {\ct}, and is given as a function of the coefficients of the same particle with an orientation angle fixed to $\theta = 0^{o}$, {\ctpl}, for the same flow conditions. The fit parameters, $\mathcal{\zeta}_i$, with $i$ being the constant number, are listed in table~\ref{table:torquecoefficient-zero}.}\label{table:torquecoefficient-summary}
  \end{table}

\section{Conclusions}\label{sec:conclusions}
In this study, we have expanded the understanding of hydrodynamic forces and torque on axi-symmetric, non-spherical rod-like particles, focusing on the complex interactions between such particles and a wall-bounded linear shear flow. The inclusion of the wall effect, a significant improvement from previous models that considered only uniform flows~\citep{Zastawny2012c,Ouchene2016,Sanjeevi2018,Frohlich2020,Sanjeevi2022}, and unbounded shear flows~\citep{Cui2018,Cheron2024}, reveals a significant change in the hydrodynamic forces and torque, compared to a uniform flow configuration. Through  direct numerical simulations, we systematically explore the variations in drag, lift, and torque coefficients for particles with varying aspect ratios, various orientation angles between the particle and the main local flow direction, various distances to the wall, and various particle Reynolds numbers, thereby addressing the limitations of existing correlations that predominantly model interactions in uniform flow conditions.\\

In the direct numerical simulations carried out in this work, the orientation angle between the main axis of the particle and the main fluid velocity direction is varied in the range $\theta = 0^{o}$ to $\theta = 180^{o}$, the particle Reynolds number in the range \Rep = 0.1 to 200, the aspect ratio of the particle in the range $\alpha = 2.5$ to $10$, and finally, the dimensionless distance between the centre of the particle and the wall, from $\tilde{L} = b$ to $\tilde{L} = \infty$. For this configuration, $b$ is the shortest distance between the centre of mass of the particle and the surface, also referred to as semi-minor axis, and $\infty$ represents the configuration when the particle is outside the boundary layer. 

Our results show that the drag coefficient for non-spherical particles, especially those with higher aspect ratios, significantly increases when the particles are closer to the wall. Moreover, the increase in the drag coefficient, which is inversely proportional to the wall distance, is larger at lower particle Reynolds numbers. The orientation of the particle also significantly influences the drag coefficient, with a maximum observed at an orientation angle of $\theta = 90^{o}$. We propose a correlation that predicts the drag coefficient of a particle located in a boundary layer based on the drag force of a particle in a uniform flow, which provides a good accuracy; a correlation coefficient of $\mathcal{R}^2 = 0.99$, and an error median of ${\mathcal{E}} = 2.89\%$, are obtained between the prediction and our simulations.\\

To describe the influence of a wall-bounded linear shear flow on the lift coefficient experienced by a non-spherical particle, we propose an additional term to the lift coefficient in a uniform flow.
This additional contribution to the lift coefficient is more significant for low particle Reynolds numbers and smaller distances to the wall.
The orientation angle also plays a role in the lift force, where we observe that the presence of the wall destroys the symmetry in the evolution of the lift coefficient as a function of the orientation angle between the particle and the mean flow.
The correlation we propose for predicting the lift coefficient of a particle in a wall boundary layer captures this accurately, and the comparison between the model predictions and our simulation results yield a correlation coefficient of $\mathcal{R}^2 = 0.99$, and an error median of ${\mathcal{E}} = 5.37\%$.\\

We also show that the torque is strongly affected by the change in the flow regime compared to the torque coefficient of a non-spherical particle in a uniform flow.
Specifically, at lower Reynolds numbers the torque coefficient undergoes substantial changes near the wall when compared to the cases with a locally uniform flow, with larger changes for particles with a larger aspect ratio. This regime also reveals that the evolution of the torque coefficient as a function of the orientation angle is symmetric about $\theta = 90^\circ$, following a ``sine-square'' profile. However, this symmetry decrease as the particle Reynolds number increases. At higher particle Reynolds numbers, the symmetry at $\theta = 90^\circ$ is lost. Nonetheless, given the challenges in accurately modelling the combined effects of wall proximity, particle aspect ratio, and orientation angles on the torque coefficient, we maintain a ``sine-square'' model assumption. This choice provides a good accuracy, since when comparing the model prediction and the PR-DNS results, a correlation coefficient of $\mathcal{R}^2 = 0.96$, and an error median of ${\mathcal{E}} = 11.00\%$.\\

The correlations derived in this study present a predictive model for the drag, lift, and torque coefficients of rod-like particles present in wall-bounded shear flows, which can be used in Eulerian-Lagrangian or discrete element modelling simulations. Moreover, these correlations are also applicable to particles outside the boundary layer, as the predictions of other research works have been used for the limits of the correlations put forward in this work.

\section*{Data availability statement} 
The data that support the findings of this study are reproducible
and files to regenerate the data as well as an executable to implement the correlations to predict the drag, lift, and torque coefficients are publically available in the repository 
with DOI 10.5281/zenodo.11059469 on \href{https://doi.org/10.5281/zenodo.11059469}{https://doi.org/10.5281/zenodo.11059469}.

\section*{Acknowledgments}
This research was funded by the Deutsche Forschungsgemeinschaft (DFG, German Research Foundation) - Project-ID 448292913.

\bibliographystyle{model1-num-names}

\end{document}